\documentclass[10pt,twocolumn,letterpaper]{article}

\usepackage{wacv}              % To produce the CAMERA-READY version
\usepackage{graphicx}
\usepackage{graphics}
\usepackage{amsmath}
\usepackage{amssymb}
\usepackage{booktabs}
\usepackage{xcolor}
\usepackage{adjustbox}
\usepackage{makecell}
\usepackage{subcaption}
\usepackage[accsupp]{axessibility}

\usepackage[pagebackref=true,breaklinks=true,letterpaper=true,colorlinks,bookmarks=false]{hyperref}

\usepackage[capitalize]{cleveref}
\crefname{section}{Sec.}{Secs.}
\Crefname{section}{Section}{Sections}
\Crefname{table}{Table}{Tables}
\crefname{table}{Tab.}{Tabs.}

\def\wacvPaperID{2122} % *** Enter the WACV Paper ID here
\def\confName{WACV}
\def\confYear{2024}

\begin{document}
%
%
%   def.set
%
%       seungjin@postech.ac.kr
%
%       Last updated:
%       September 19, 2004
%
%       A set of my definitions and new commands
%
%
% math bold for lower cases (a-z)
\newcommand{\ba}{{\mathbf{a}}}
\newcommand{\bb}{{\mathbf{b}}}
\newcommand{\bc}{{\mathbf{c}}}
\newcommand{\bd}{{\mathbf{d}}}
\newcommand{\bolde}{{\mathbf{e}}}
\newcommand{\boldf}{{\mathbf{f}}}
\newcommand{\bg}{{\mathbf{g}}}
\newcommand{\bh}{{\mathbf{h}}}
\newcommand{\bi}{{\mathbf{i}}}
\newcommand{\bj}{{\mathbf{j}}}
\newcommand{\bk}{{\mathbf{k}}}
\newcommand{\bl}{{\mathbf{l}}}
\newcommand{\bm}{{\mathbf{m}}}
\newcommand{\bn}{{\mathbf{n}}}
\newcommand{\bo}{{\mathbf{o}}}
\newcommand{\bp}{{\mathbf{p}}}
\newcommand{\bq}{{\mathbf{q}}}
\newcommand{\br}{{\mathbf{r}}}
\newcommand{\bs}{{\mathbf{s}}}
\newcommand{\bt}{{\mathbf{t}}}
\newcommand{\bu}{{\mathbf{u}}}
\newcommand{\bv}{{\mathbf{v}}}
\newcommand{\bw}{{\mathbf{w}}}
\newcommand{\bx}{{\mathbf{x}}}
\newcommand{\by}{{\mathbf{y}}}
\newcommand{\bz}{{\mathbf{z}}}

% math bold for upper cases (A-Z)
\newcommand{\bA}{\mathbf{A}}
\newcommand{\bB}{\mathbf{B}}
\newcommand{\bC}{\mathbf{C}}
\newcommand{\bD}{\mathbf{D}}
\newcommand{\bE}{\mathbf{E}}
\newcommand{\bF}{\mathbf{F}}
\newcommand{\bG}{\mathbf{G}}
\newcommand{\bH}{\mathbf{H}}
\newcommand{\bI}{\mathbf{I}}
\newcommand{\bJ}{\mathbf{J}}
\newcommand{\bK}{\mathbf{K}}
\newcommand{\bL}{\mathbf{L}}
\newcommand{\bM}{\mathbf{M}}
\newcommand{\bN}{\mathbf{N}}
\newcommand{\bO}{\mathbf{O}}
\newcommand{\bP}{\mathbf{P}}
\newcommand{\bQ}{\mathbf{Q}}
\newcommand{\bR}{\mathbf{R}}
\newcommand{\bS}{\mathbf{S}}
\newcommand{\bT}{\mathbf{T}}
\newcommand{\bU}{\mathbf{U}}
\newcommand{\bV}{\mathbf{V}}
\newcommand{\bW}{\mathbf{W}}
\newcommand{\bX}{\mathbf{X}}
\newcommand{\bY}{\mathbf{Y}}
\newcommand{\bZ}{\mathbf{Z}}

% calligraphic
\newcommand{\calA}{{\mathcal{A}}}
\newcommand{\calB}{{\mathcal{B}}}
\newcommand{\calC}{{\mathcal{C}}}
\newcommand{\calD}{{\mathcal{D}}}
\newcommand{\calE}{{\mathcal{E}}}
\newcommand{\calF}{{\mathcal{F}}}
\newcommand{\calG}{{\mathcal{G}}}
\newcommand{\calH}{{\mathcal{H}}}
\newcommand{\calI}{{\mathcal{I}}}
\newcommand{\calJ}{{\mathcal{J}}}
\newcommand{\calK}{{\mathcal{K}}}
\newcommand{\calL}{{\mathcal{L}}}
\newcommand{\calM}{{\mathcal{M}}}
\newcommand{\calN}{{\mathcal{N}}}
\newcommand{\calO}{{\mathcal{O}}}
\newcommand{\calP}{{\mathcal{P}}}
\newcommand{\calQ}{{\mathcal{Q}}}
\newcommand{\calR}{{\mathcal{R}}}
\newcommand{\calS}{{\mathcal{S}}}
\newcommand{\calT}{{\mathcal{T}}}
\newcommand{\calU}{{\mathcal{U}}}
\newcommand{\calV}{{\mathcal{V}}}
\newcommand{\calW}{{\mathcal{W}}}
\newcommand{\calX}{{\mathcal{X}}}
\newcommand{\calY}{{\mathcal{Y}}}
\newcommand{\calZ}{{\mathcal{Z}}}
\newcommand{\calbX}{\mbox{\boldmath $\mathcal{X}$}}
\newcommand{\calbY}{\mbox{\boldmath $\mathcal{Y}$}}

\newcommand{\bcalA}{\mbox{\boldmath $\calA$}}
\newcommand{\bcalB}{\mbox{\boldmath $\calB$}}
\newcommand{\bcalC}{\mbox{\boldmath $\calC$}}
\newcommand{\bcalD}{\mbox{\boldmath $\calD$}}
\newcommand{\bcalE}{\mbox{\boldmath $\calE$}}
\newcommand{\bcalF}{\mbox{\boldmath $\calF$}}
\newcommand{\bcalG}{\mbox{\boldmath $\calG$}}
\newcommand{\bcalH}{\mbox{\boldmath $\calH$}}
\newcommand{\bcalI}{\mbox{\boldmath $\calI$}}
\newcommand{\bcalJ}{\mbox{\boldmath $\calJ$}}
\newcommand{\bcalK}{\mbox{\boldmath $\calK$}}
\newcommand{\bcalL}{\mbox{\boldmath $\calL$}}
\newcommand{\bcalM}{\mbox{\boldmath $\calM$}}
\newcommand{\bcalN}{\mbox{\boldmath $\calN$}}
\newcommand{\bcalO}{\mbox{\boldmath $\calO$}}
\newcommand{\bcalP}{\mbox{\boldmath $\calP$}}
\newcommand{\bcalQ}{\mbox{\boldmath $\calQ$}}
\newcommand{\bcalR}{\mbox{\boldmath $\calR$}}
\newcommand{\bcalS}{\mbox{\boldmath $\calS$}}
\newcommand{\bcalT}{\mbox{\boldmath $\calT$}}
\newcommand{\bcalU}{\mbox{\boldmath $\calU$}}
\newcommand{\bcalV}{\mbox{\boldmath $\calV$}}
\newcommand{\bcalW}{\mbox{\boldmath $\calW$}}
\newcommand{\bcalX}{\mbox{\boldmath $\calX$}}
\newcommand{\bcalY}{\mbox{\boldmath $\calY$}}
\newcommand{\bcalZ}{\mbox{\boldmath $\calZ$}}

\newcommand{\sfA}{\mbox{$\mathsf A$}}
\newcommand{\sfB}{\mbox{$\mathsf B$}}
\newcommand{\sfC}{\mbox{$\mathsf C$}}
\newcommand{\sfD}{\mbox{$\mathsf D$}}
\newcommand{\sfE}{\mbox{$\mathsf E$}}
\newcommand{\sfF}{\mbox{$\mathsf F$}}
\newcommand{\sfG}{\mbox{$\mathsf G$}}
\newcommand{\sfH}{\mbox{$\mathsf H$}}
\newcommand{\sfI}{\mbox{$\mathsf I$}}
\newcommand{\sfJ}{\mbox{$\mathsf J$}}
\newcommand{\sfK}{\mbox{$\mathsf K$}}
\newcommand{\sfL}{\mbox{$\mathsf L$}}
\newcommand{\sfM}{\mbox{$\mathsf M$}}
\newcommand{\sfN}{\mbox{$\mathsf N$}}
\newcommand{\sfO}{\mbox{$\mathsf O$}}
\newcommand{\sfP}{\mbox{$\mathsf P$}}
\newcommand{\sfQ}{\mbox{$\mathsf Q$}}
\newcommand{\sfR}{\mbox{$\mathsf R$}}
\newcommand{\sfS}{\mbox{$\mathsf S$}}
\newcommand{\sfT}{\mbox{$\mathsf T$}}
\newcommand{\sfU}{\mbox{$\mathsf U$}}
\newcommand{\sfV}{\mbox{$\mathsf V$}}
\newcommand{\sfW}{\mbox{$\mathsf W$}}
\newcommand{\sfX}{\mbox{$\mathsf X$}}
\newcommand{\sfY}{\mbox{$\mathsf Y$}}
\newcommand{\sfZ}{\mbox{$\mathsf Z$}}

% math bold for lower cases (Greek letters)
\newcommand{\balpha}{\mbox{\boldmath $\alpha$}}
\newcommand{\bbeta}{\mbox{\boldmath $\beta$}}
\newcommand{\bgamma}{\mbox{\boldmath $\gamma$}}
\newcommand{\bdelta}{\mbox{\boldmath $\delta$}}
\newcommand{\bepsilon}{\mbox{\boldmath $\epsilon$}}
\newcommand{\bvarepsilon}{\mbox{\boldmath $\varepsilon$}}
\newcommand{\bzeta}{\mbox{\boldmath $\zeta$}}
\newcommand{\boldeta}{\mbox{\boldmath $\eta$}}
\newcommand{\btheta}{\mbox{\boldmath $\theta$}}
\newcommand{\bvartheta}{\mbox{\boldmath $\vartheta$}}
\newcommand{\biota}{\mbox{\boldmath $\iota$}}
\newcommand{\bkappa}{\mbox{\boldmath $\kappa$}}
\newcommand{\blambda}{\mbox{\boldmath $\lambda$}}
\newcommand{\bmu}{\mbox{\boldmath $\mu$}}
\newcommand{\bnu}{\mbox{\boldmath $\nu$}}
\newcommand{\bxi}{\mbox{\boldmath $\xi$}}
\newcommand{\bpi}{\mbox{\boldmath $\pi$}}
\newcommand{\bvarpi}{\mbox{\boldmath $\varpi$}}
\newcommand{\brho}{\mbox{\boldmath $\rho$}}
\newcommand{\bvarrho}{\mbox{\boldmath $\varrho$}}
\newcommand{\bsigma}{\mbox{\boldmath $\sigma$}}
\newcommand{\bvarsigma}{\mbox{\boldmath $\varsigma$}}
\newcommand{\btau}{\mbox{\boldmath $\tau$}}
\newcommand{\bupsilon}{\mbox{\boldmath $\upsilon$}}
\newcommand{\bphi}{\mbox{\boldmath $\phi$}}
\newcommand{\bvarphi}{\mbox{\boldmath $\varphi$}}
\newcommand{\bchi}{\mbox{\boldmath $\chi$}}
\newcommand{\bpsi}{\mbox{\boldmath $\psi$}}
\newcommand{\bomega}{\mbox{\boldmath $\omega$}}

% math bold for upper cases (Greek Letters)
\newcommand{\bGamma}{\mbox{\boldmath $\Gamma$}}
\newcommand{\bDelta}{\mbox{\boldmath $\Delta$}}
\newcommand{\bTheta}{\mbox{\boldmath $\Theta$}}
\newcommand{\bLambda}{\mbox{\boldmath $\Lambda$}}
\newcommand{\bXi}{\mbox{\boldmath $\Xi$}}
\newcommand{\bPi}{\mbox{\boldmath $\Pi$}}
\newcommand{\bSigma}{\mbox{\boldmath $\Sigma$}}
\newcommand{\bUpsilon}{\mbox{\boldmath $\Upsilon$}}
\newcommand{\bPhi}{\mbox{\boldmath $\Phi$}}
\newcommand{\bPsi}{\mbox{\boldmath $\Psi$}}
\newcommand{\bOmega}{\mbox{\boldmath $\Omega$}}

% vector notation for lower cases (a-z)
\newcommand{\veca}{{\vec{\ba}}}
\newcommand{\vecb}{{\vec{\bb}}}
\newcommand{\vecc}{{\vec{\bc}}}
\newcommand{\vecd}{{\vec{\bd}}}
\newcommand{\vece}{{\vec{\bolde}}}
\newcommand{\vecf}{{\vec{\boldf}}}
\newcommand{\vecg}{{\vec{\bg}}}
\newcommand{\vech}{{\vec{\bh}}}
\newcommand{\veci}{{\vec{\bi}}}
\newcommand{\vecj}{{\vec{\bj}}}
\newcommand{\veck}{{\vec{\bk}}}
\newcommand{\vecl}{{\vec{\bl}}}
\newcommand{\vecm}{{\vec{\bm}}}
\newcommand{\vecn}{{\vec{\bn}}}
\newcommand{\veco}{{\vec{\bo}}}
\newcommand{\vecp}{{\vec{\bp}}}
\newcommand{\vecq}{{\vec{\bq}}}
\newcommand{\vecr}{{\vec{\br}}}
\newcommand{\vecs}{{\vec{\bs}}}
\newcommand{\vect}{{\vec{\bt}}}
\newcommand{\vecu}{{\vec{\bu}}}
\newcommand{\vecv}{{\vec{\bv}}}
\newcommand{\vecw}{{\vec{\bw}}}
\newcommand{\vecx}{{\vec{\bx}}}
\newcommand{\vecy}{{\vec{\by}}}
\newcommand{\vecz}{{\vec{\bz}}}

\newcommand{\vecxi}{{\vec{\bxi}}}
\newcommand{\vecphi}{{\vec{\bphi}}}
\newcommand{\vecvarphi}{{\vec{\bvarphi}}}
\newcommand{\vecbeta}{{\vec{\bbeta}}}
\newcommand{\vecdelta}{{\vec{\bdelta}}}
\newcommand{\vectheta}{{\vec{\btheta}}}

% set of numbers
\newcommand{\Real}{\mathbb R}
\newcommand{\Complex}{\mathbb C}
\newcommand{\Natural}{\mathbb N}
\newcommand{\Integer}{\mathbb Z}

% color
%\definecolor{orange}{rgb}{1,0.3,0}
%\definecolor{copper}{rgb}{1,.62,.40}

% ets
\newcommand{\bone}{\mbox{\boldmath $1$}}
\newcommand{\bzero}{\mbox{\boldmath $0$}}
\newcommand{\0}{{\bf 0}}

\newcommand{\be}{\begin{eqnarray}}
\newcommand{\ee}{\end{eqnarray}}
\newcommand{\bee}{\begin{eqnarray*}}
\newcommand{\eee}{\end{eqnarray*}}

\newcommand{\matrixb}{\left[ \begin{array}}
\newcommand{\matrixe}{\end{array} \right]}

\newcommand{\argmax}{\operatornamewithlimits{\arg \max}}
\newcommand{\argmin}{\operatornamewithlimits{\arg \min}}

\newcommand{\mean}[1]{\left \langle #1 \right \rangle}
\newcommand{\ave}{\mathbb E}
\newcommand{\E}{\mathbb E}
\newcommand{\empha}[1]{{\color{red} \bf #1}}
\newcommand{\fracpartial}[2]{\frac{\partial #1}{\partial  #2}}
\newcommand{\incomplete}[1]{\textcolor{red}{#1}}

\def\doublespace{\renewcommand{\baselinestretch}{2}\large\normalsize}
\def\singlespace{\renewcommand{\baselinestretch}{1}\large\normalsize}
\def\onehalfspace{\renewcommand{\baselinestretch}{1.5}\large\normalsize}
\def\onequaterspace{\renewcommand{\baselinestretch}{1.3}\large\normalsize}
\def\threequaterspace{\renewcommand{\baselinestretch}{1.7}\large\normalsize}
\def\smallspace{\renewcommand{\baselinestretch}{-.9}\large\normalsize}
\def\tinyspace{\renewcommand{\baselinestretch}{-.7}\large\normalsize}

\newcommand{\tr} { \textrm{tr} }
\newcommand{\re} { \textrm{re} }
\newcommand{\im} { \textrm{im} }
\newcommand{\diag} { \textrm{diag} }
\newcommand{\ddiag} { \textrm{ddiag} }
\newcommand{\off} { \textrm{off} }
\newcommand{\vectxt} { \textrm{vec} }

\newcommand{\lla}{\left\langle}
\newcommand{\rra}{\right\rangle}
\newcommand{\llbr}{\left\lbrack}
\newcommand{\rrbr}{\right\rbrack}
\newcommand{\llb}{\left\lbrace}
\newcommand{\rrb}{\right\rbrace}

% Extra commands==================

\newcommand{\RR}{I\!\!R} %real numbers
\newcommand{\Nat}{I\!\!N} %natural numbers
\newcommand{\CC}{I\!\!\!\!C} %complex numbers

\newcommand{\Tref}[1]{Table~\ref{#1}}
\newcommand{\Eref}[1]{Eq.~(\ref{#1})}
\newcommand{\Fref}[1]{Fig.~\ref{#1}}
\newcommand{\FCref}[1]{Chapter.~\ref{#1}}
\newcommand{\Sref}[1]{Sec.~\ref{#1}}
\newcommand{\Aref}[1]{Algo.~\ref{#1}}

\def\eg{\emph{e.g.}}
\def\Eg{\emph{E.g. }}
\def\etal{\emph{et al. }}
\def\ie{\emph{i.e. }}

%%%%%%%%% TITLE
\title{Blurry Video Compression

A Trade-off between Visual Enhancement and Data Compression}

\author{Dawit Mureja Argaw \\ KAIST \and Junsik Kim \\ Harvard University \and In So Kweon \\ KAIST}

\maketitle

%%%%%%%%% ABSTRACT
\begin{abstract}
Existing video compression (VC) methods primarily aim to reduce the spatial and temporal redundancies between consecutive frames in a video while preserving its quality. In this regard, previous works have achieved remarkable results on videos acquired under specific settings such as instant (known) exposure time and shutter speed which often result in sharp videos. However, when these methods are evaluated on videos captured under different temporal priors, which lead to degradations like motion blur and low frame rate, they fail to maintain the quality of the contents. In this work, we tackle the VC problem in a general scenario where a given video can be blurry due to predefined camera settings or dynamics in the scene. By exploiting the natural trade-off between visual enhancement and data compression, we formulate VC as a min-max optimization problem and propose an effective framework and training strategy to tackle the problem. Extensive experimental results on several benchmark datasets confirm the effectiveness of our method compared to several state-of-the-art VC approaches. 

\end{abstract}
\vspace{-2mm}
\section{Introduction}
\label{sec:intro}
\vspace{-1mm}
Video compression (VC) methods primarily aim to jointly compress the \textit{motion} estimated between consecutive frames and the \textit{residual} computed between the reconstructed frames and their original counterpart. 
In this regard, existing VC approaches~\cite{yang2020learninga,yang2020learningb,lu2019dvc,wu2018video,Agustsson_2020_CVPR,Lin_2020_CVPR,sun2020high,hu2020improving,lu2020content,liu2020conditional, Hu_2021_CVPR,li2021deep,mentzer2022vct,mentzer2022neural} have achieved remarkable results on videos acquired under specific settings such as instant (known) exposure time and shutter speed that often result in distinctively sharp videos with sufficiently high frame rate. However, when these methods are evaluated on videos captured under different temporal priors, such as slow shutter speed, long exposure time, and fast-moving objects, which lead to degradations like motion blur and low frame rate, they perform very poorly and fail to preserve the input video quality. This is mainly because motion and residual information cannot be precisely computed and compressed between frames with degraded contents and relatively large temporal distances, and hence existing approaches cannot effectively generalize to complicated real-world scenarios. Furthermore, this limitation cannot be trivially solved by retraining previous works with blurry footage owing to the nature of their problem formulation.

With the recent progress in deep network-based motion deblurring~\cite{wang2019edvr,gao2019dynamic,nah2017deep,su2017deep,tao2018srndeblur}, a straightforward approach to tackling the task at hand would be to cascade off-the-shelf deblurring and video compression models. However, using cascade models results in sub-optimal performance. First, the pixel errors introduced in the deblurring stage would propagate to the compression stage, thus degrading the overall performance. Second, due to the arbitrariness of the artifacts introduced in the deblurring stage, since each frame is processed independently, temporal smoothness can not be ensured. As a result, the decoded video will suffer from flickering artifacts. The alternative strategy of reverse cascading (compression followed by deblurring) also suffers from the same limitations. 

An attempt to address these limitations by deploying a na\"ive end-to-end optimization scheme on cascade models, unexpectedly, results in an even worse performance. This is mainly due to the inherent trade-off between visual enhancement and data compression when performing the joint task. Intuitively speaking, a blurry video can be regarded as the pseudo-compressed version of its sharp equivalent. On the other hand, a blind attempt to enhance the visual quality of the given blurry video before compression would in turn have a decompressing effect, \ie~it increases the amount of bits required to encode the given data. 

We have observed that an end-to-end training of cascade models with the standard rate-distortion optimization \cite{lu2019dvc,yang2020learninga,yang2020learningb} fails to exploit this trade-off and makes the cascade models converge to a \textit{bad local minima} \cite{liang2018adding,sohl2019eliminating,kawaguchi2020elimination}, where the deblurring network collapses to an identity function in deblurring + compression (\texttt{D} + \texttt{C}) cascade models, whereas the compression network incurs a heavy encoding cost in compression + deblurring (\texttt{C} + \texttt{D}) cascade models. 

This paper tackles the video compression problem in a situation, where a video may contain blurry regions due to a predefined camera setting or dynamics in a scene. Inspired by the aforementioned trade-off, we formulate the task at hand as a min-max optimization problem, \ie~we adaptively maximize the visual quality of a given video while simultaneously minimizing the number of bits needed to encode it. To this end, we present a single, end-to-end trainable network with interrelated visual enhancement and compression modules. To mitigate erroneous motion estimation due to blur corrupted pixels or large displacements, we propose a context-aware flow refinement mechanism. Finally, we devise an effective training strategy to tailor each component in our network toward optimal compression performance. We show the versatility of our approach by demonstrating that \textit{one} well-trained model can be used for 1) generating sharp, yet compressed videos from raw blurry videos and 2) compressing videos while preserving the input prior, \ie~ either sharp or blur prior.
\vspace{-1mm}
\paragraph{Contributions.} To summarize, our contributions are two-fold: 
(\textbf{1}) To the best of our knowledge, this is the first work that studies neural compression for blurry videos. Besides proper problem formulation, our work introduces an effective framework to tackle the problem. 
(\textbf{2}) We comprehensively analyze our work and competing baselines in several settings and confirm the merit of the proposed approach.
\section{Related works}
Early VC algorithms such as H.264 \cite{wiegand2003overview} and H.265 \cite{sullivan2012overview} achieve highly efficient compression performance, however, they are based on manually designed modules. In the following years, several works \cite{xu2018reducing,liu2018one,dai2017convolutional,li2019densenet,li2019deep} have attempted to substitute some of the components in traditional codecs with DNN-based methods, yet none of these works are end-to-end trainable. Meanwhile, Wu \etal~\cite{wu2018video} formulated the VC problem as an interpolation between compressed images and trained a deep network in an end-to-end manner. However, their model performance is below the widely used video coding standards. Lu \etal~\cite{lu2019dvc} mimicked the pipeline in conventional methods and proposed the first deep VC framework that has achieved state-of-the-art performance. Most follow-up works focused on improving compression performance using: multiple reference frames \cite{yang2020learninga,Lin_2020_CVPR,yang2020learningb,Rippel_2019_ICCV}, hierarchical compression \cite{yang2020learninga}, recurrent models \cite{yang2020learningb}, adaptive blurring \cite{Agustsson_2020_CVPR}, feature-level compression~\cite{djelouah2019vidcodec,Feng_2020_CVPR_Workshops,Hu_2021_CVPR}, contextual video coding~\cite{li2021deep}, transformer-based architecture~\cite{mentzer2022vct} and GAN losses to add realism to decoded videos~\cite{mentzer2022neural,yang2021perceptual}. However, most previous works are prone to motion artifacts, such as blur and low frame rate, and fail to preserve image quality when performing compression under these settings. Our work tackles this problem via joint formulation of visual enhancement and data compression.
\section{Methodology}
\begin{figure*}[!t]
    \centering
    \includegraphics[width=1\linewidth, trim={1.17cm 0.3cm 1.23cm 0.9cm},clip]{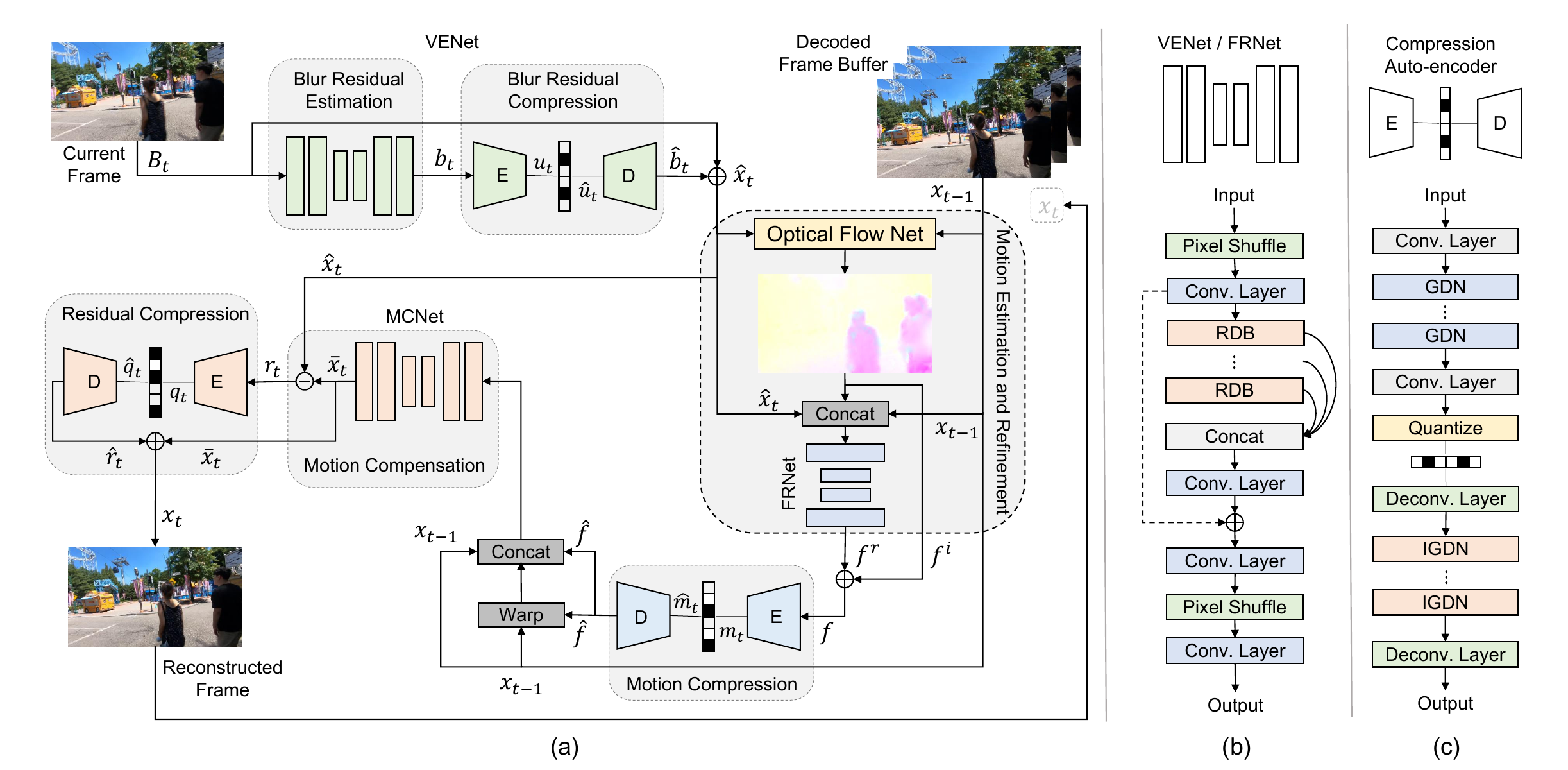}
    \vspace{-4mm}
    \caption{\small (a) Overview of our proposed framework. (b) The backbone structure for blur and flow residual estimation networks. (c) Network architecture for the compression autoencoder.}
    \vspace{-5mm}
    \label{fig:model}
\end{figure*}

\textbf{Problem Formulation.} Given a sequence of blurred inputs $\{B_t, B_{t+1}, \ldots, B_{t+n}\}$, we aim to output a visually sharp yet compressed video $\{x_t, x_{t+1}, \ldots, x_{t+n}\}$. Let $\calQ$ represent a metric to measure the visual quality of frames and $\calS$ represent the number of bits required to encode a given data. We formulate the joint blur reduction and video compression task as a min-max problem by optimizing a single model $\calF$ with parameters $\Theta$ in an end-to-end manner with a compression loss $\calL$ as follows,
\vspace{-0.5mm}
\begin{equation}
\begin{aligned}
& \underset{\mathbb{U} \subseteq \Theta}{\text{min}} \hspace{0.9mm} \underset{\mathbb{V} \subseteq \Theta}{\text{max}}
& &  \sum \calL \Big(\calF_{\Theta} \left. \Big(\big\{B_t\big\}_{t=1}^{n} | x_{t-1}\Big) \right\vert \big\{X_t\big\}_{t=1}^{n} \Big)\\
& \text{subject to}
& & \calQ\big(\big\{x_t\big\}_{t=1}^{n}\big) \gg \calQ\big(\big\{B_t\big\}_{t=1}^{n}\big)\\
& & & \calS\big(\big\{x_t\big\}_{t=1}^{n}\big) \ll \calS\big(\big\{B_t\big\}_{t=1}^{n}\big)\\
\end{aligned}
\end{equation}

where $x_{t-1}$ denotes a keyframe encoded using an image compression method similar to `I-frame' in traditional codecs~\cite{vanne2012comparative,ohm2012comparison,mukherjee2013latest}, $\{X_t, X_{t+1}, \ldots, X_{t+n}\}$ represents the ground truth (GT)  sequence, $\mathbb{V}$ denotes a subset of model parameters that contribute towards maximizing the quality of the input blurry frames with respect to their sharp counterparts and $\mathbb{U}$ represents a subset of model parameters associated with quantizing motion and residual information to minimize temporal redundancy between frames.

Since the visual enhancement process increases the Bpp of a video, optimizing $\mathbb{V}$ has a decompressing effect that contradicts the compression objective $\calL$. Therefore, joint training of \textit{blur reduction} and \textit{video compression} becomes a min-max problem. As mentioned previously, na\"ive joint training does not work since the two tasks counter each other. In this work, we propose a framework that optimizes these two contradicting objectives. The overview of our proposed framework is depicted in \Fref{fig:model}.

\subsection{Visual Enhancement}
\label{sec:step1}
To reduce the blur of the given input $B_t$, we design a visual enhancement network (\texttt{VENet}) which has two interrelated components. The first part estimates an additive vector representation $b_t$, which we refer to as \textit{blur residual}, to offset the blur from $B_t$. If the given input is a sharp frame, then $b_t$ is approximately a zero vector of the same size as the given input. The second part of the \texttt{VENet}, on the other hand, compresses the blur residual information. The compressed blur residual $\hat{b}_t$ will then be added to $B_t$ to output a visually enhanced frame $\hat{x}_t$ that will be used in the next steps (see \Fref{fig:model}a). The main motivation behind such a design choice comes in twofold. First, unlike blind deblurring approaches \cite{wang2019edvr,gao2019dynamic,nah2017deep,su2017deep,tao2018srndeblur} that directly output a deblurred frame, our \texttt{VENet} enables us to simultaneously optimize the quality and size of the enhanced frame. Second, as our network is trained in an end-to-end manner, the \texttt{VENet} learns to maximize the quality of the input frame for optimal video compression. 

To estimate the blur residual $b_t$ conditioned on the input $B_t$, we used a variation of residual dense network \cite{zhang2018residual} as a backbone network. As shown in  \Fref{fig:model}b, the backbone network consists of a pixel shuffle layer, convolutional layers, residual dense blocks (RDB) \cite{zhang2018residual}, and a sub-pixel convolution (pixel reshuffle) layer \cite{shi2016real}. In order to compress the estimated blur residual information $b_t$, we adopted an auto-encoder style network \cite{balle2016end,lu2019dvc,yang2020learninga}. As depicted in \Fref{fig:model}c, the blur residual is fed into a series of convolution (deconvolution) and nonlinear transform layers. 
Given a blur residual $b_t$ of size $M \times N \times 3$, the encoder generates a blur residual representation $u_t$ of size $M/16 \times N/16 \times 128$. Then $u_t$ is quantized to $\hat{u}_t$. We used the \textit{factorized} entropy model \cite{balle2018variational} for quantization. The decoder inputs the quantized representation and reconstructs the blur residual information $\hat{b}_t$. The enhanced frame $\hat{x}_t$ is then obtained by adding $\hat{b}_t$ to the blurred input $B_t$, \ie~$\hat{x}_t = B_t + \hat{b}_t$.

\begin{figure*}[!t]
\begin{center}
\setlength{\tabcolsep}{0.3pt}
\renewcommand{\arraystretch}{0.25}
\resizebox{1.0\linewidth}{!}{%
\begin{tabular}{cccccc}
          \tiny{$x_{t-1}$} & \tiny{$B_t$} & \tiny{$B_t$ crop} & \tiny{$\calE$} & \tiny{$f^i_{t \rightarrow t-1}$} & \tiny{$f_{t \rightarrow t-1}$} \\
          \includegraphics[width=0.1\linewidth]{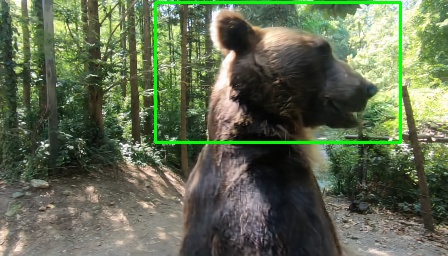}&
         \includegraphics[width=0.1\linewidth]{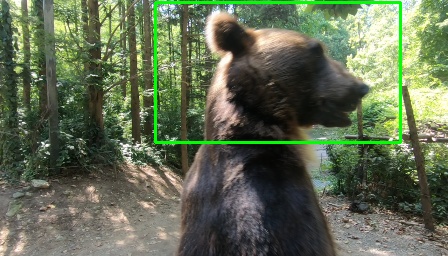}&
         \includegraphics[width=0.1\linewidth]{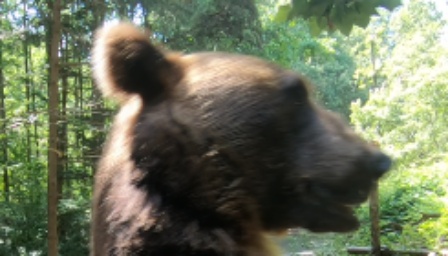}&
         \includegraphics[width=0.1\linewidth]{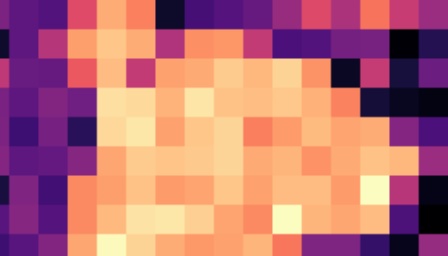}&
         \includegraphics[width=0.1\linewidth]{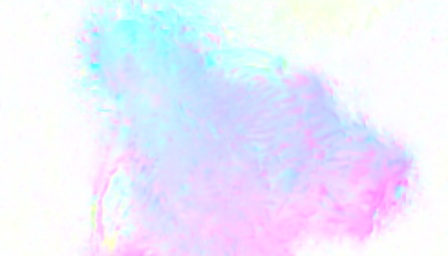}&
         \includegraphics[width=0.1\linewidth]{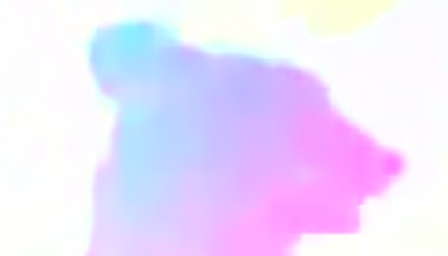}
\end{tabular}}
\vspace{-4mm}
\end{center}
\caption{\small Visualization of motion refinement. $\calE$, $f^i_{t \rightarrow t-1}$ and $f_{t \rightarrow t-1}$ denote the attention map, initial flow, and refined flow, respectively.}
\vspace{-4mm}
\label{fig:qual_vis}
\end{figure*}

\subsection{Motion Estimation and Compression}
\label{sec:step2}
\vspace{-0.5mm}
To reduce temporal redundancy in the given video, we first estimate the motion between the current enhanced frame $\hat{x}_t$ and the previous reconstructed frame $x_{t-1}$. We employ a pre-trained optical flow network \cite{ranjan2017optical} to predict an initial flow $f^i_{t \rightarrow t-1}$. Most existing video compression works fine-tune the pre-trained flow network using a warp loss (\Eref{eqn: warp}) for better motion estimation. However, we argue that such a modification has two important limitations. First, it assumes that $\hat{x}_t$ and  $x_{t-1}$ are absolute references, and thus does not attend to potential motion artifacts in the inputs. This in turn leads to erroneous results (see \Sref{sec:ablation}). Second, as the receptive field of the flow network is fixed, it is challenging to adaptively handle large motions. To address these limitations, we propose a flow refinement network (\texttt{FRNet}) and an attention-based loss function for effective motion estimation and compression.

As shown in \Fref{fig:model}a, \texttt{FRNet} inputs the initial flow $f^i_{t \rightarrow t-1}$, $\hat{x}_t$ and  $x_{t-1}$ and outputs a residual flow $f^r_{t \rightarrow t-1}$ which will be added to the initial flow to generate a refined flow (\Eref{eqn: frnet}, \Eref{eqn: fr}). We used a residual dense architecture (see \Fref{fig:model}b) with three RDBs to generate $f^r_{t \rightarrow t-1}$. 

The refined flow information $f_{t \rightarrow t-1}$ is then encoded, quantized, and reconstructed as $\hat{f}_{t \rightarrow t-1}$ using a flow auto-encoder network (see \Fref{fig:model}a).
\begin{equation}
    \calL_\texttt{warp} = \sum \big\|\hat{x}_t - \calW_b(x_{t-1}, \hat{f}^i_{t \rightarrow t-1})\big\|_2
    \label{eqn: warp}
\end{equation}
\vspace{-4mm}
\begin{equation}
    f^r_{t \rightarrow t-1} = \texttt{FRNet}(\hat{x}_t~||~x_{t-1}~||~f^i_{t \rightarrow t-1})
    \label{eqn: frnet}
\end{equation}
\vspace{-3mm}
\begin{equation}
     f_{t \rightarrow t-1} =  f^i_{t \rightarrow t-1} +  f^r_{t \rightarrow t-1}
    \label{eqn: fr}
\end{equation}
where $\calW_b$ denotes a back-warping layer,  $ \hat{f}^i_{t \rightarrow t-1}$ denotes the reconstructed initial flow and $||$ stands for channel-wise concatenation. 

To tailor the flow refinement process for the task at hand, we designed a context-aware training loss that enforces \texttt{FRNet} to attend to the visually enhanced regions. Specifically, we generate an attention map by scoring the different regions of $\hat{x}_t$ according to their degree of enhancement (with respect to $B_t$) so that the motion refinement stage knows which regions to particularly focus on (see \Fref{fig:qual_vis}). To achieve this, we first compute the error map $\calE$ which is defined as the mean squared error between $\hat{x}_t$ and the corresponding GT frame $X_t$, \ie~$\|\hat{x}_t -X_t\|_2$. $\calE$ is a 2D tensor of size $M \times N$, where values are averaged across channel. To avoid a noisy map, we further applied an average pooling layer of kernel size $k$ and stride $k$ and assign each pixel in the error map with the corresponding average value of its neighborhood, \ie~we segmented $\calE$ into $\frac{M\cdot N}{k^2}$ regions of size $k \times k$. Then we rank and label each region with integer values $\{v, \ldots, v\frac{M\cdot N}{k^2}\}$, where $v$ is a constant to ensure increased variance in the error map distribution. 
\vspace{-0.5mm}
\begin{equation}
\calE= \texttt{rank}_{[v, v\frac{M\cdot N}{k^2}]}\Big[\texttt{AvgPool}_{(k,k)} \big(\|\hat{x}_t -X_t\|_2 \big)\Big]
\label{eqn: e_map}
\vspace{-0.5mm}
\end{equation}
The higher values in $\calE$ represent segments that still have motion artifacts while the lower values represent regions that are enhanced (or that were sharp initially). By using $\calE$ as an attention weight, we propose a new loss, which we refer to as \textit{context-aware loss} ($\calL_\texttt{CaL}$), for informed motion refinement and compression as shown in \Eref{eqn: cal}. $\calE$ is min-max scaled  between 0 and 1 (\Eref{eqn:scale}) in order not to disrupt the natural trade-off between distortion and bit-rate during training (refer to \Sref{sec:training}).

\begin{equation}
    w = \frac{\calE - \min(\calE)}{\max(\calE) - \min(\calE)}
    \label{eqn:scale}
\end{equation}

\begin{equation}
    \calL_\texttt{CaL} = \sum \Big\|w \cdot \big[\hat{x}_t - \calW_b(x_{t-1}, \hat{f}_{t \rightarrow t-1})\big]\Big\|_2
    \label{eqn: cal}
\end{equation}

\vspace{-2mm}
\subsection{Residual Compression}
\label{sec:step3}

Given the refined flow $\hat{f}_{t \rightarrow t-1}$, we back-warp the reference frame $x_{t-1}$ to reconstruct the current frame. To compensate for the artifacts in the warped frame, we further process it using a motion compensation network (\texttt{MCNet}). As depicted in \Fref{fig:model}a, \texttt{MCNet} inputs the warped frame, the reference frame $x_{t-1}$ and the motion vector $\hat{f}_{t \rightarrow t-1}$ and outputs a motion compensated frame $\overline{x}_t$ which is expected to be as close as the enhanced frame $\hat{x}_t$ (\Eref{eqn: mc}). We used a residual-UNet architecture like the one used in \cite{lu2019dvc,yang2020learninga} for \texttt{MCNet}. 

\begin{equation}
    \overline{x}_t = \texttt{MCNet}\big(\calW_b(x_{t-1}, \hat{f}_{t \rightarrow t-1})~||~x_{t-1}~||~\hat{f}_{t \rightarrow t-1}\big)
    \label{eqn: mc}
\end{equation}
Finally, the residual between the enhanced raw frame $\hat{x}_t$ and the motion compensated frame $\overline{x}_t$, \ie~$r_t = \hat{x}_t - \overline{x}_t$, is compressed using a residual encoder-decoder network. Like the blur residual and motion compression, the residual information $r_t$ is first encoded to a latent representation $q_t$, then quantized to $\hat{q_t}$ and finally decoded to $\hat{r}_t$  (see \Fref{fig:model}a). The reconstructed residual information $\hat{r}_t$ is added to the motion compensated frame $\overline{x}_t$ to obtain the compressed frame $x_t$, \ie~$x_t = \overline{x}_t + \hat{r}_t$.

\subsection{Training Strategy}
\label{sec:training}

The goal of our video compression framework is to minimize the number of bits used for encoding a given video frame $B_t$, while simultaneously enhancing its quality with respect to its sharp counterpart $X_t$ and reducing the distortion between the enhanced frame $\hat{x}_t$ and the reconstructed frame $x_t$. Therefore, we formulate the optimization problem as follows,
\vspace{-2mm}
\begin{equation}
    \calL = \lambda_e E + \lambda_d D + R
    \label{eqn: opt}
    \vspace{-1mm}
\end{equation}
where $\lambda_e$ and $ \lambda_d$ are hyperparameters to control the three-way trade-off between the enhancement $E$, distortion $D$ and bit-rate $R$. For the visual enhancement part, we jointly optimize the number of encoding bits for the quantized blur residual $\hat{u}_t$ and the $\ell1$ photometric loss between the enhanced frame $\hat{x}_t$ and the corresponding GT frame $X_t$ as shown in \Eref{eqn:loss_venet}. Note that we also included the $\ell1$ photometric loss between $B_t + b_t$ and $X_t$ so that the blur residual $b_t$ auto-encoder will not collapse into a \textit{bad local minima} where $\hat{b}_t = 0$, \ie~$B_t = \hat{x}_t$.
\vspace{-1mm}
\begin{equation}
    \calL_\texttt{VENet} = \lambda_e \Big[\sum  \big\|X_t - (B_t+b_t)\big\|_1 + \big\|X_t - \hat{x}_t\big\|_1\Big] + R (\hat{u}_t)
    \label{eqn:loss_venet}
    \vspace{-1mm}
\end{equation}
where $R(\cdot)$ denotes the number of bits used for encoding the representations. We used the density model of \cite{balle2016end} to estimate $R$. $\lambda_e$ is defined as $\lambda_e = s\lambda_d$, where $s$ is a step decay parameter to maintain the trade-off between the visual enhancement and compression as the training progresses.

Following previous works \cite{lu2019dvc,yang2020learninga,Lin_2020_CVPR,yang2020learningb}, we used a progressive scheme to train the different components in the compression part. First, the motion estimation and compression step in \Sref{sec:step2} is trained by optimizing the proposed context-aware loss $\calL_\texttt{CaL}$ and the bit-rate for encoding the quantized motion vector $\hat{m}_t$ (\Eref{eqn:loss_motion}). The optical flow network \cite{ranjan2017optical} used to obtain the initial flow $f^i_{t \rightarrow t-1}$ is initialized with pre-trained weights and remains unchanged. Then, the motion compensation network in \Sref{sec:step3} is added into the training using the loss function defined in \Eref{eqn:loss_mcnet}.
\begin{equation}
    \calL_\texttt{M} = \lambda_d \calL_\texttt{CaL} + R (\hat{m}_t) 
    \label{eqn:loss_motion}
\end{equation}
\begin{equation}
    \calL_\texttt{MCNet} = \lambda_d \sum \big\|\overline{x}_t - \hat{x}_t \big\|_2  + R (\hat{m}_t)
    \label{eqn:loss_mcnet}
\end{equation}
Finally, the distortion between the reconstructed frame $x_t$ and the enhanced frame $\hat{x}_t$ is optimized using the loss function formulated in \Eref{eqn:loss_dis}, where $D(x,y)$ is defined as a distortion metrics. We used the mean square error (MSE), \ie~$D(x,y) = \texttt{MSE}(x,y)$ when optimizing for PSNR and $D(x,y) = 1-\texttt{MS-SSIM}(x,y)$ when optimizing for MS-SSIM. This is in accordance with the experiment protocol in previous works \cite{lu2019dvc,yang2020learninga,yang2020learningb}. The total training loss for end-to-end optimization of the whole network is summarized in \Eref{eqn:loss_total}.
\begin{equation}
    \calL_\texttt{D} = \lambda_d D(x_t,\hat{x}_t) + R (\hat{m}_t) + R (\hat{q}_t)
    \label{eqn:loss_dis}
\end{equation}
\vspace{-3mm}
\begin{equation}
    \calL_\texttt{total}(i) = \begin{cases} 
      \calL_\texttt{VENet}   & i \leq a \\
       \calL_\texttt{VENet} + \calL_\texttt{M} & a < i \leq b \\
       \calL_\texttt{VENet} + \calL_\texttt{MCNet} & b < i \leq  c \\
       \calL_\texttt{VENet} + \calL_\texttt{D} & c < i \leq \max_\texttt{iter}
   \end{cases}
   \label{eqn:loss_total}
\end{equation}
where $a$, $b$, $c$ are different iteration steps and $\max_\texttt{iter}$ is the maximum iteration during training. 
\section{Experiments}
\subsection{Experimental Setting}
\paragraph{Datasets.} Existing works use Vimeo-90k dataset~\cite{xue2019video}, which contains $89,800$ clips each with $7$ sharp frames, for a model training. However, as this dataset cannot be applied to train a model in a blurry scenario, we follow the common practice in computer vision research and generate a blur dataset by synthesizing low-frame-rate videos from a sharp high-frame-rate sequence  \cite{shen2020blurry,zhang2020video}. To simulate a frame $B_t$ taken by a low-frame-rate camera, we average $n$ consecutive frames taken by a $240$ fps camera. 
\vspace{-1mm}
\begin{equation*}
    B_t = \frac{1}{n}\sum_{j = \kappa t - \frac{n}{2}}^{j = \kappa t + \frac{n}{2}} I_j
\vspace{-1mm}
\end{equation*}

where $I_j$ is the $j$-th high-frame-rate latent image and the parameter $\kappa$ determines the frame rate of the acquired frames. As $n$ is related to the degree of blur \cite{brooks2019learning}, we experiment with different values of $n \in \{5,7,9\}$. To take the shutter closing time into account during video acquisition \cite{zhang2020video}, we discard $m$ number of consecutive frames in between before synthesizing the next frame $B_{t+1}$, where $m \in \{5,3,1\}$. We set $\kappa = m+n = 10$ to downsample the $240$ fps to $24$ fps, which is a common fps setting for commodity cameras. In this manner, we create blurry videos captured under different exposure settings. 

We apply this scheme to Adobe240~\cite{su2017deep}, GOPRO~\cite{nah2017deep} and REDS \cite{Nah_2019_CVPR_Workshops_REDS}  datasets which provide high-frame-rate videos with a resolution of $1280 \times 720$. Most of the videos in these datasets, however, have less than a thousand frames which makes it challenging to synthesize enough and diverse training set. Hence, instead of training separately on each dataset, we used a total of $325$ videos ($100$ from Adobe240, $25$ from GOPRO, $200$ from REDS) and built a training dataset that approximately has $32,500$ clips each with $7$ frames. The remaining non-overlapping videos in each dataset ($30$ from Adobe240, $8$ from GOPRO, and $70$ from REDS) are used for testing. We also make use of other benchmark datasets such as UVG \cite{mercat2020uvg} and MCL-JCV~\cite{wang2016mcl} for generalization experiments.
\vspace{-3.5mm}
\paragraph{Implementation Details.} Following the previous works \cite{lu2019dvc,yang2020learninga,yang2020learningb}, we train eight models by setting $\lambda_d = 256, 512, 1024, 2048$ when optimizing for PSNR and  $\lambda_d = 8, 16, 32, 64$ when optimizing for MS-SSIM. For compressing the reference `I-frame', we use BPG \cite{bellard2015bpg} and Lee \etal~\cite{lee2018context} in our PSNR and MS-SSIM models, respectively. For each model, an Adam optimizer \cite{kingma2014adam} with an initial learning rate $1e-4$ and momentum $0.9$ is used. We set the different iteration steps $a$, $b$, $c$ and $\max_\texttt{iter}$ to $1e+5$, $2e+5$, $3e+5$ and $1e+6$, respectively. The learning rate is decayed by a factor of 10 at $5e+5$ and $8e+5$ iterations. The step decay parameter $s$ for $\lambda_e$ is initialized with $1$ and reduced by $0.25$ every $2e+5$ iterations. We use a mini-batch size of 4 by randomly cropping images of size $256 \times 256$ during training. Our network is implemented using Tensorflow \cite{abadi2016tensorflow}.
\vspace{-3.5mm}
\paragraph{Evaluation Metrics.} We measure the distortion between the reconstructed frame $x_t$ and the GT frame $X_t$ on PSNR and MS-SSIM metrics with respect to the number of bits for encoding the representations $m_t$ and $q_t$. Bits per pixel (Bpp) is used to represent the required bits for each pixel in the current frame.

%%%%%%%%%%%%%%%%%%%%%%%%%%%%%%%%%%%%%%%%%%%%%%%%%%%%%%%%%%%%%%%%%%%%%%%%%%%%
\begin{figure*}[!t]
    \centering
    \begin{subfigure}[]{0.24\textwidth}
    \centering
    \includegraphics[width = \textwidth, trim={0.4cm 0.25cm 0.35cm 0.4cm},clip]{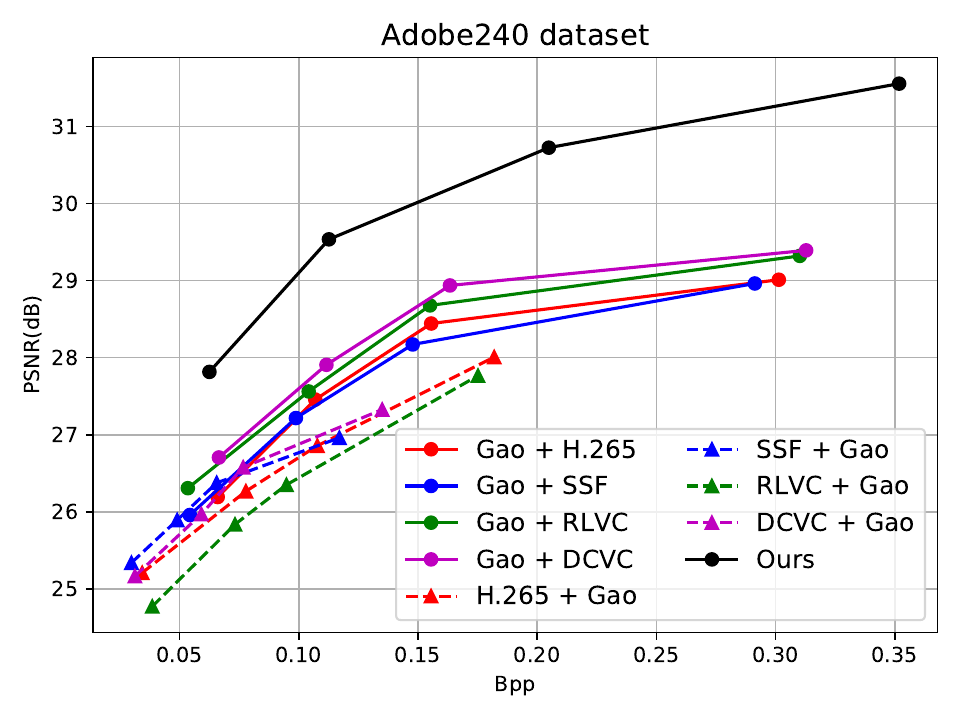}
    \end{subfigure}
    \hfill
    \begin{subfigure}[]{0.24\textwidth}
    \centering
    \includegraphics[width = \textwidth, trim={0.4cm 0.25cm 0.35cm 0.4cm},clip]{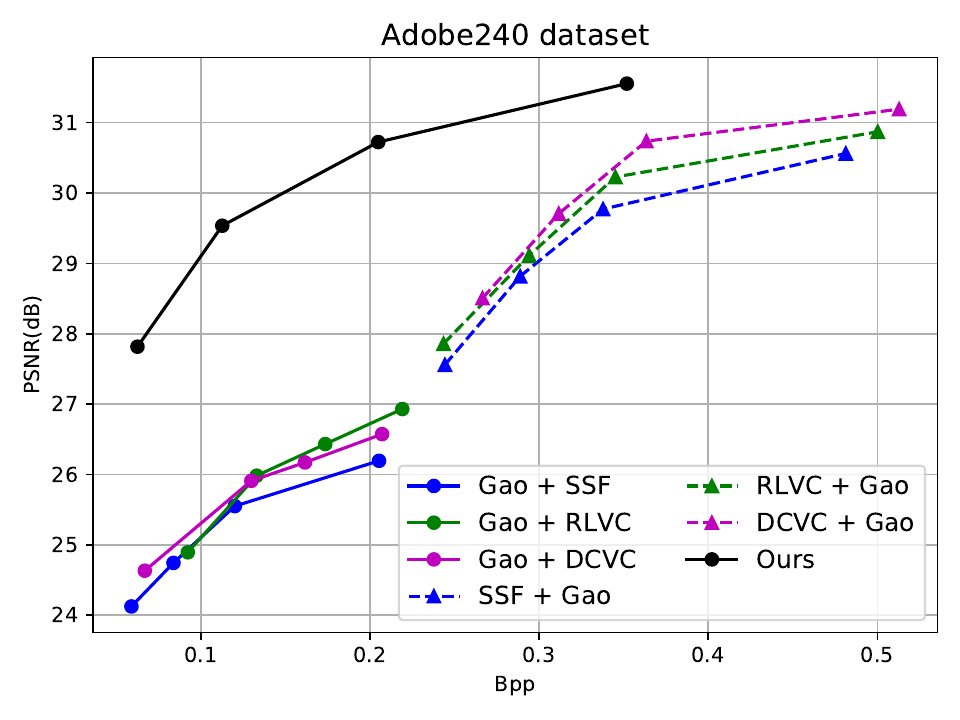}
    \end{subfigure}
    \hfill
    \begin{subfigure}[]{0.24\textwidth}
    \centering
    \includegraphics[width = \textwidth, trim={0.4cm 0.25cm 0.35cm 0.4cm},clip]{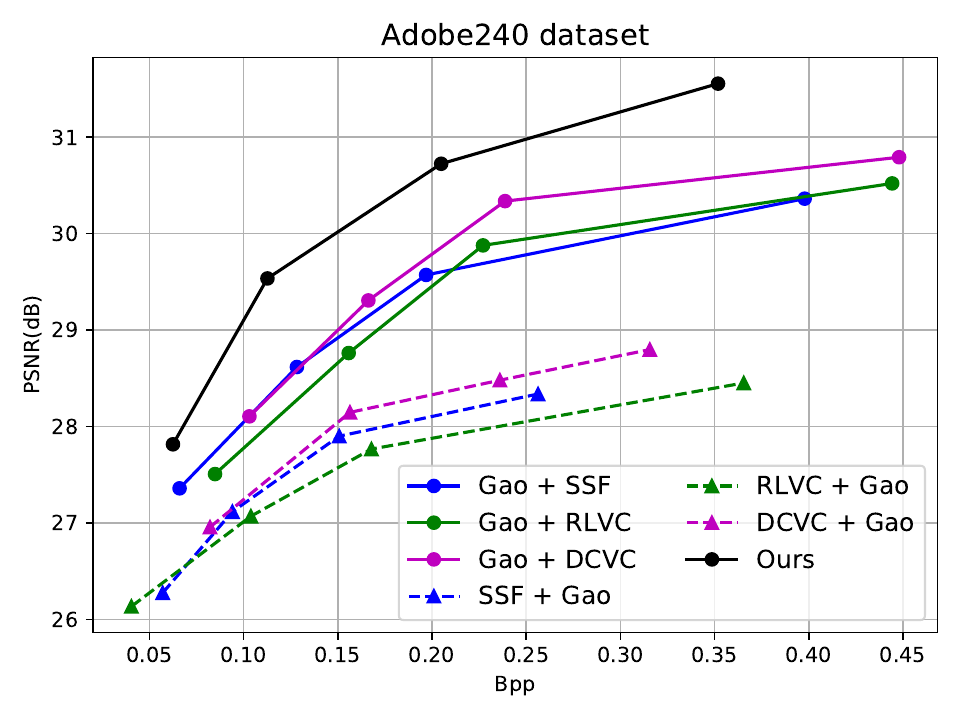}
    \end{subfigure}
    \hfill
    \begin{subfigure}[]{0.24\textwidth}
    \centering
    \includegraphics[width = \textwidth, trim={0.4cm 0.25cm 0.35cm 0.4cm},clip]{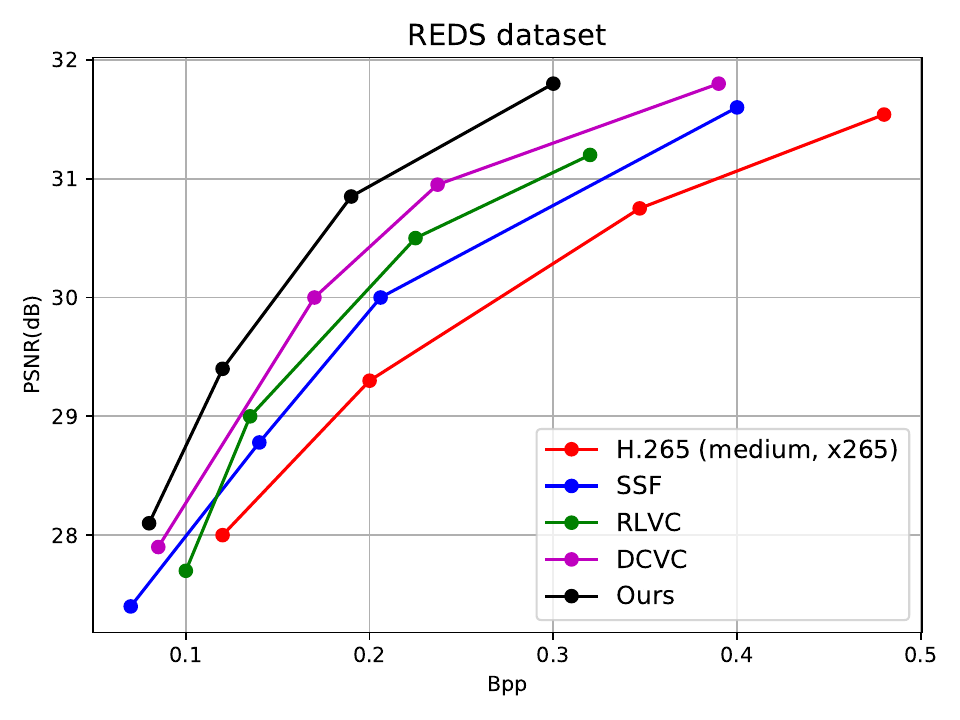}
    \end{subfigure}
    \hfill
    \vspace{1mm}
    \begin{subfigure}[]{0.24\textwidth}
    \centering
    \includegraphics[width = \textwidth, trim={0.4cm 0.25cm 0.45cm 0.4cm},clip]{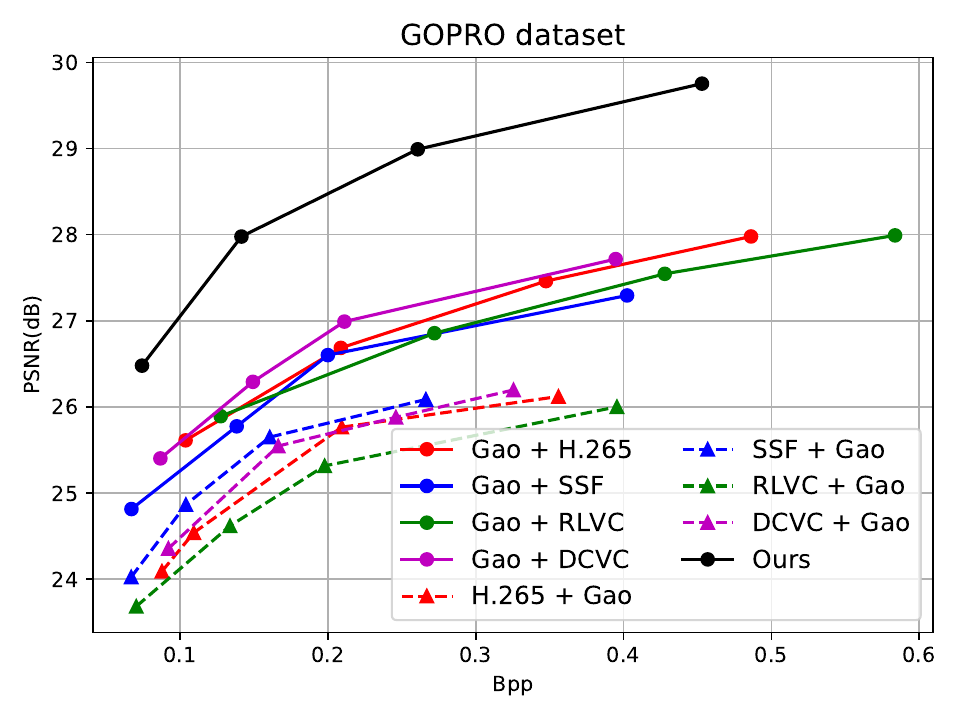}
    \end{subfigure}
    \hfill
    \begin{subfigure}[]{0.24\textwidth}
    \centering
    \includegraphics[width = \textwidth, trim={0.4cm 0.25cm 0.35cm 0.4cm},clip]{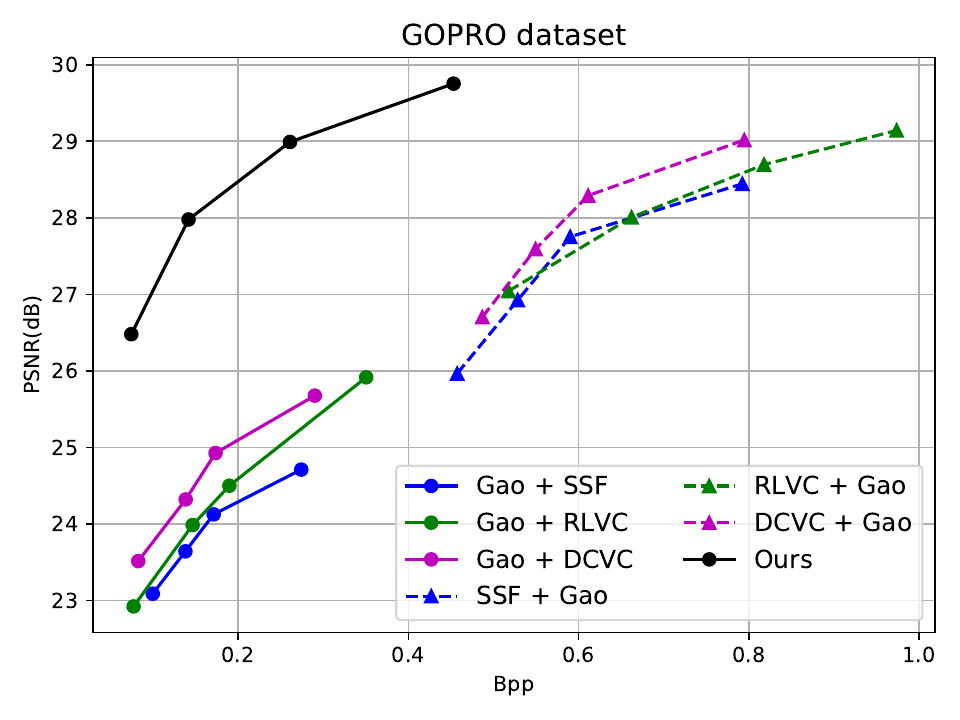}
    \end{subfigure}
    \hfill
    \begin{subfigure}[]{0.24\textwidth}
    \centering
    \includegraphics[width = \textwidth, trim={0.4cm 0.25cm 0.35cm 0.4cm},clip]{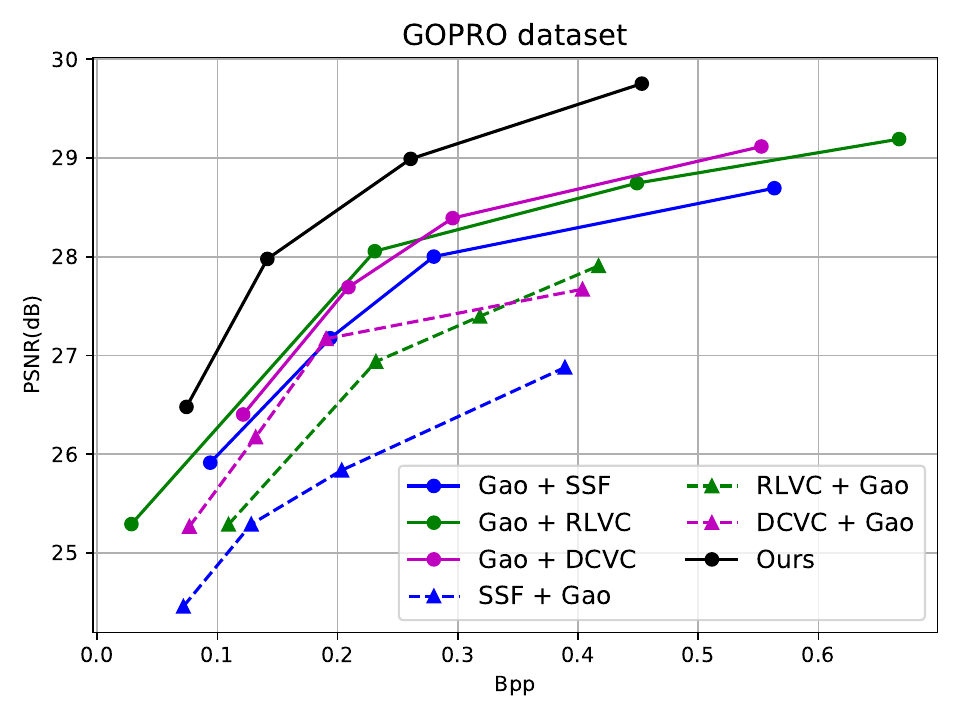}
    \end{subfigure}
    \hfill
    \begin{subfigure}[]{0.24\textwidth}
    \centering
    \includegraphics[width = \textwidth, trim={0.4cm 0.25cm 0.35cm 0.4cm},clip]{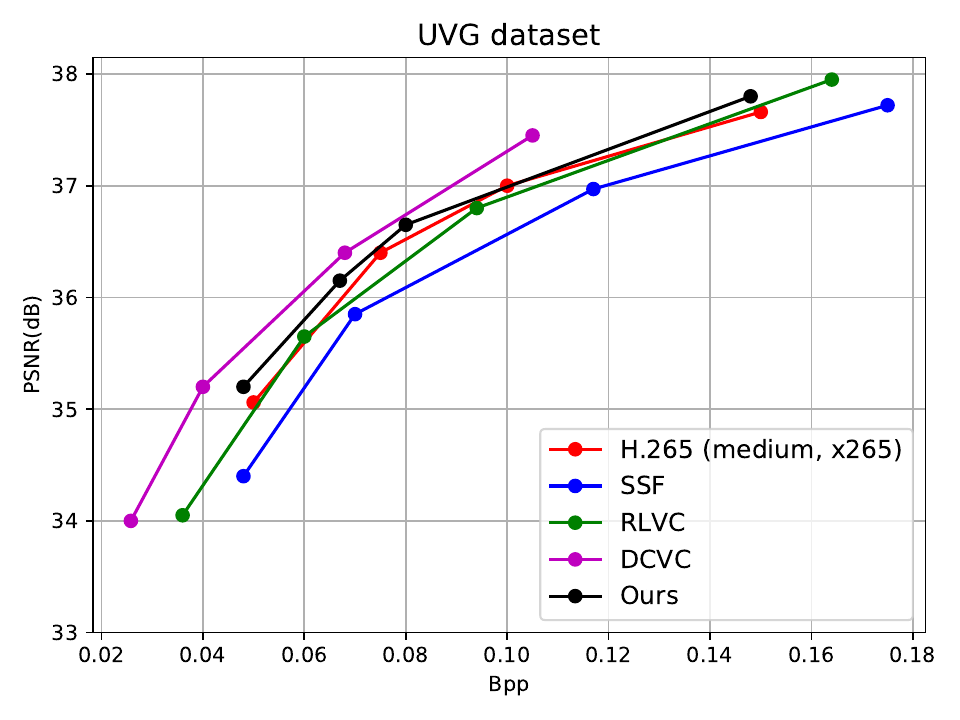}
    \end{subfigure}
    \hfill
     \vspace{1mm}
    \begin{subfigure}[]{0.24\textwidth}
    \centering
    \includegraphics[width = \textwidth, trim={0.4cm 0.25cm 0.35cm 0.4cm},clip]{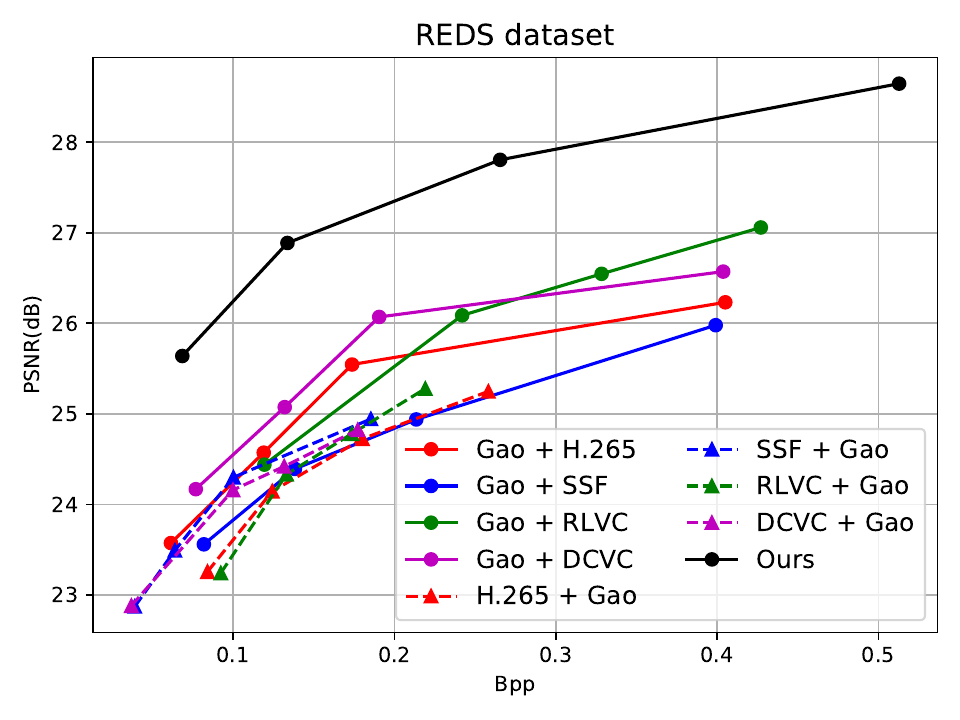}
    \caption{\textbf{Off-the-shelf}}
    \end{subfigure}
    \hfill
    \begin{subfigure}[]{0.24\textwidth}
    \centering
    \includegraphics[width = \textwidth, trim={0.4cm 0.25cm 0.35cm 0.4cm},clip]{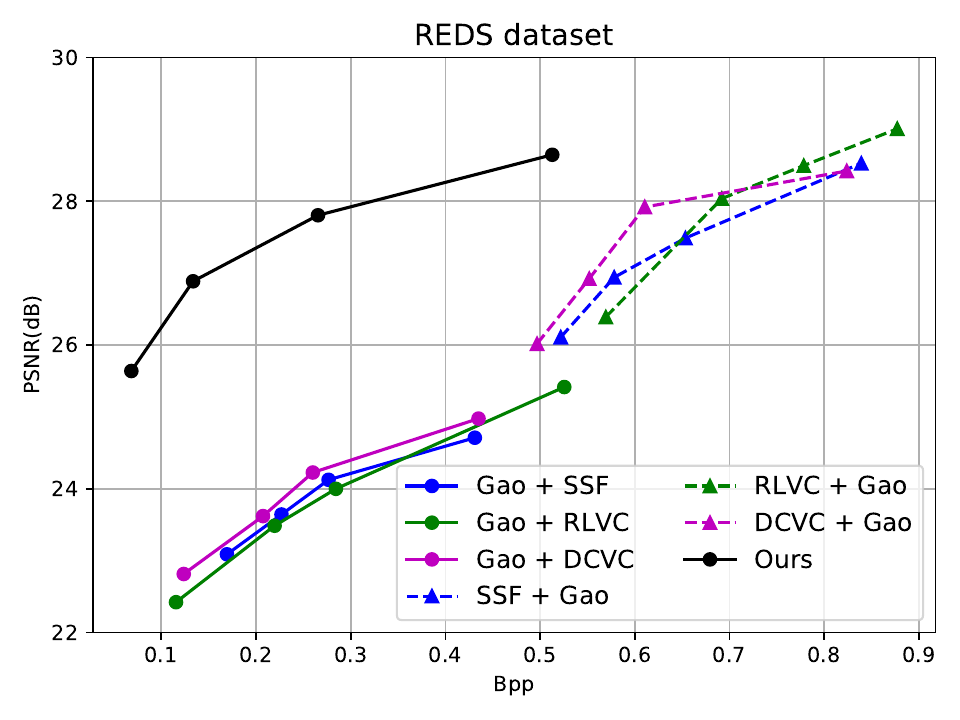}
    \caption{\textbf{Joint training}}
    \end{subfigure}
    \hfill
    \begin{subfigure}[]{0.24\textwidth}
    \centering
    \includegraphics[width = \textwidth, trim={0.4cm 0.25cm 0.35cm 0.4cm},clip]{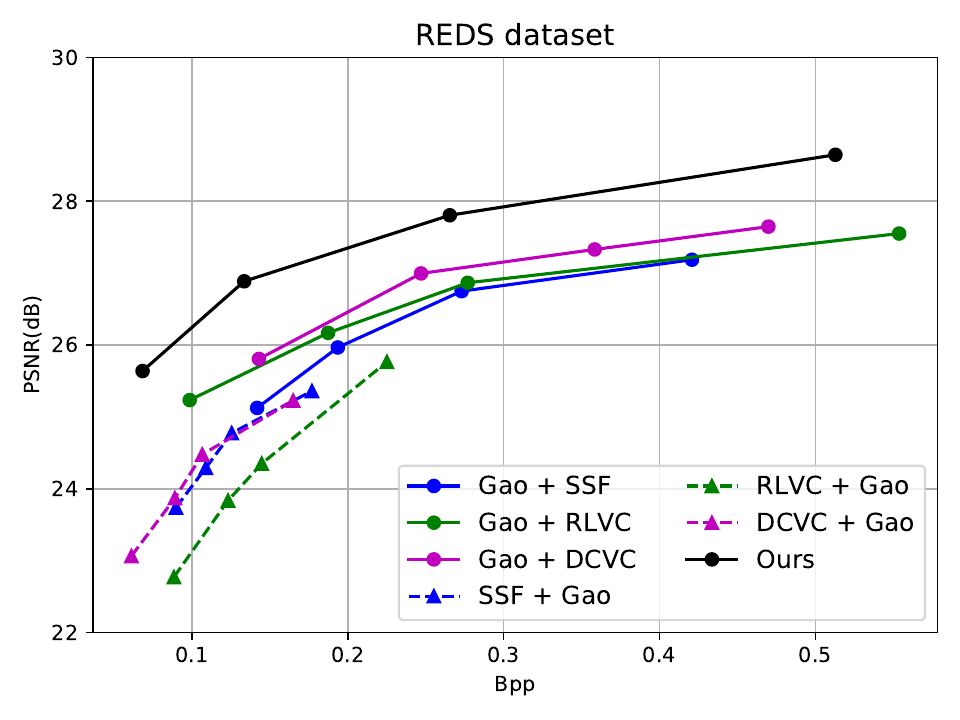}
    \caption{\textbf{Intermediate supervision}}
    \end{subfigure}
    \hfill
    \begin{subfigure}[]{0.24\textwidth}
    \centering
    \includegraphics[width = \textwidth, trim={0.4cm 0.25cm 0.35cm 0.4cm},clip]{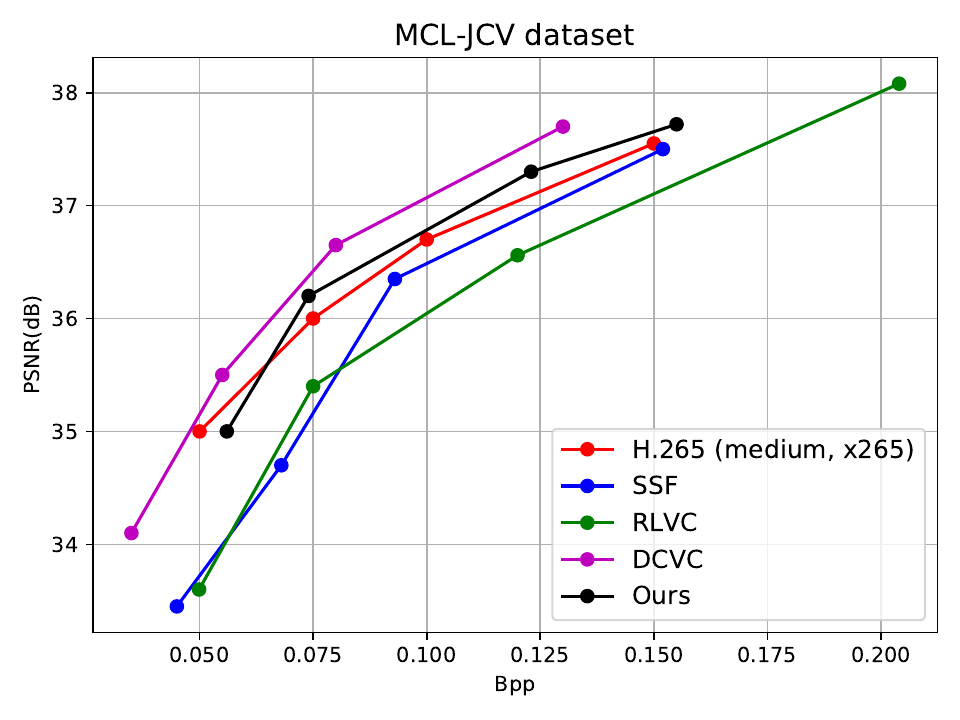}
    \caption{\textbf{Preserving the input prior}}
    \end{subfigure}
    \hfill
    \vspace{-1mm}
    \caption{\small (\textbf{a}) - (\textbf{c}) Rate-distortion performance comparison between our method and (\textbf{a}) \textit{\textbf{off-the-shelf}} cascade models. (\textbf{b}) cascade models optimized with the \textit{\textbf{joint training}} scheme ($\calL_{\texttt{Joint}}$). (\textbf{c}) cascade models optimized with \textit{\textbf{intermediate supervision}} scheme ($\calL_\texttt{Int}$). (\textbf{d}) Rate-distortion performance comparison on preserving the blur (\textit{top}) and sharp (\textit{middle} and \textit{bottom}) priors.}
    \vspace{-3mm}
    \label{fig:exp_results}
\end{figure*}
%%%%%%%%%%%%%%%%%%%%%%%%%%%%%%%%%%%%%%%%%%%%%%%%%%%%%%%%%%%%%%%%%%%%%%%%

\subsection{Experimental Results}
We compare our work with traditional video codecs, \ie~H.265~\cite{sullivan2012overview}, and the latest learned methods for which open-source implementations are available, \ie~SSF~\cite{Agustsson_2020_CVPR}, RLVC~\cite{yang2020learninga} and DCVC~\cite{li2021deep}. However, naive VC models are prone to motion artifacts and fail to preserve image quality. Hence, to compare our model with competitive baselines, we implement a cascade approach by using the state-of-the-art deblurring networks, \ie~Gao \cite{gao2019dynamic}. We create two types of cascade models, \ie~deblurring + compression (\texttt{D} + \texttt{C}) and compression + deblurring (\texttt{C} + \texttt{D}).

For each cascade model, we establish three types of baselines: (i) \textbf{\textit{Off-the-shelf}}, where pretrained models from the respective tasks are successively used for the joint task. (ii) \textbf{\textit{Joint training}}, where cascade models are jointly trained from scratch using the standard rate-distortion loss~\cite{yang2020learningb}, \ie~$\calL_\texttt{Joint} = \lambda D_\texttt{Joint} + R$, where $D_\texttt{Joint}$ denotes the losses computed with respect to the GT frame $X_t$ at different stages of training. For \texttt{D} + \texttt{C} models, $D_\texttt{Joint}$ is computed with respect to the output of the compression network (without explicit supervision to the deblurring network) and for \texttt{C} + \texttt{D} models, $D_\texttt{Joint}$ is computed with respect to the output of the deblurring network. $R$ represents the rate optimization  for motion and residual representations. (iii) \textbf{\textit{Intermediate supervision}}, where we use additional intermediate supervision between the cascaded components. For \texttt{D} + \texttt{C} models, an $\ell1$ photometric loss is used to train the deblurring network, and the compression network is simultaneously optimized with respect to the deblurred output. The total training loss is defined in \Eref{eq : optim2}.
\begin{equation}
\calL_\texttt{Int}^{\texttt{D} + \texttt{C}} = \lambda_d \big\|X_t - \hat{x}_t\big\|_1 + \lambda_c D_\texttt{Int} + R
\label{eq : optim2}
\end{equation}
where $D_\texttt{Int}$ denotes the distortion losses computed with respect to the deblurred output $\hat{x}_t$. $\lambda_d$ and $\lambda_c$ represent hyperparameters for the deblurring and compression networks, respectively. Note that the loss in \Eref{eq : optim2} is analogous to the proposed training strategy in \Sref{sec:training}. For \texttt{C} + \texttt{D} models, the equivalent training scheme would be to first optimize the compression network with the blurry input and enhance the decoded output using the deblurring network as shown in \Eref{eq : optim2_2}. 
\vspace{-2mm}
\begin{equation}
\calL_\texttt{Int}^{ \texttt{C} + \texttt{D}} =  \lambda_c D_\texttt{Int} + \lambda_d \big\|X_t - x_t\big\|_1 + R
\label{eq : optim2_2}
\vspace{-2mm}
\end{equation}
where $D_\texttt{Int}$ represents the distortion losses computed with respect to the blurry input $B_t$. We use the official code of each compression and deblurring model to implement the cascade baselines. For fair evaluation, each cascade model is trained and evaluated using the same settings and datasets as our method. We implement H.265~\cite{sullivan2012overview} in a \textit{medium} setting using $\times265$ encoder. 

%%%%%%%%%%%%%%%%%%%%%%%%%%%%%%%%%%%%%%%%%%%%%
\begin{figure*}[!ht]
\tiny
\begin{center}
\setlength{\tabcolsep}{0.4pt}
\renewcommand{\arraystretch}{0.7}
\resizebox{1.0\linewidth}{!}{%
\begin{tabular}{cccccc}
         \tiny{\texttt{Bpp}} & \tiny{\texttt{0.07101}} & \tiny{\texttt{0.0644}} & \tiny{\texttt{Bpp}} & \tiny{\texttt{0.0176}} & \tiny{\texttt{0.0164}}\\
          \includegraphics[width=0.1\linewidth]{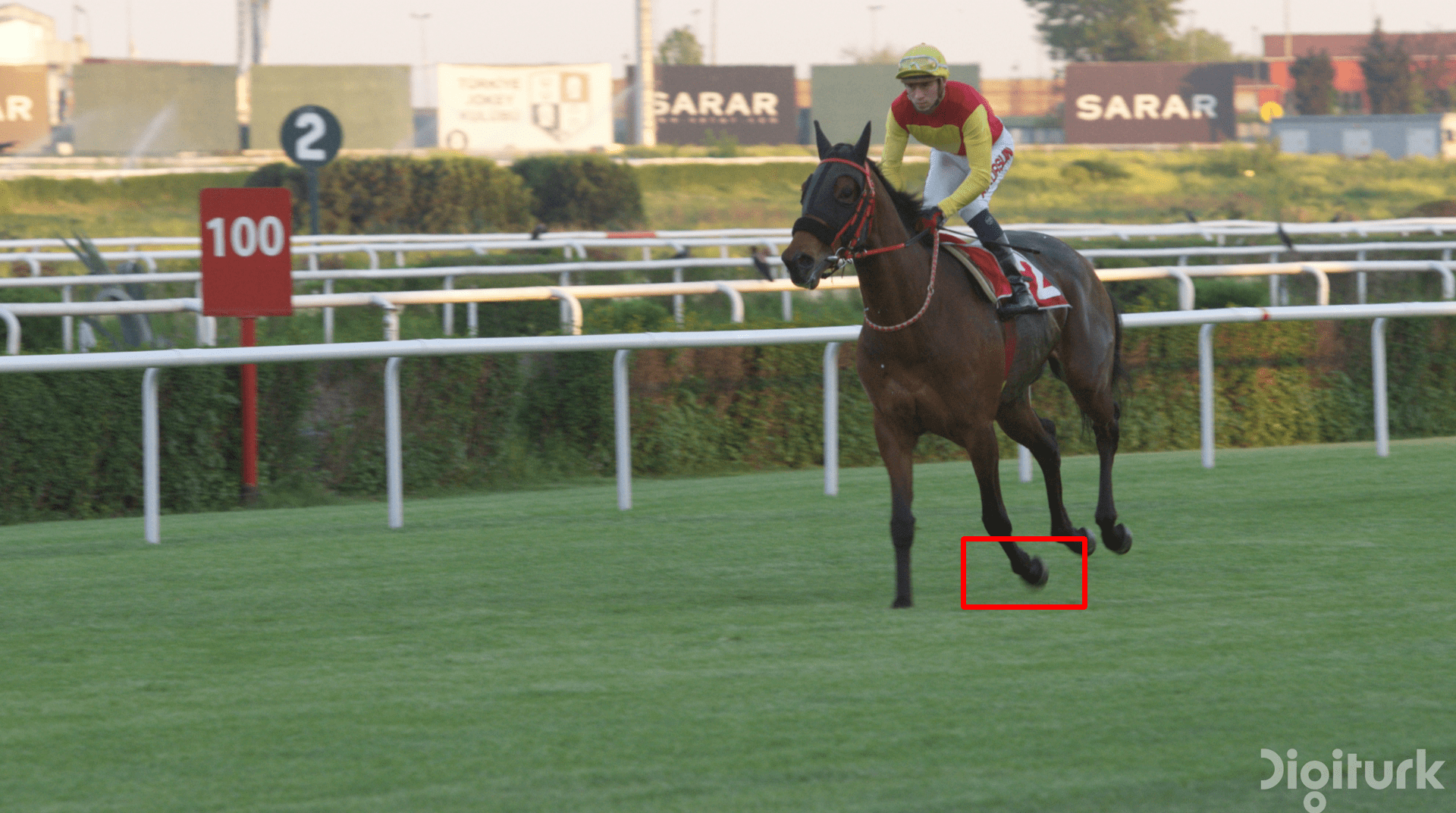}&
          \includegraphics[width=0.1\linewidth]{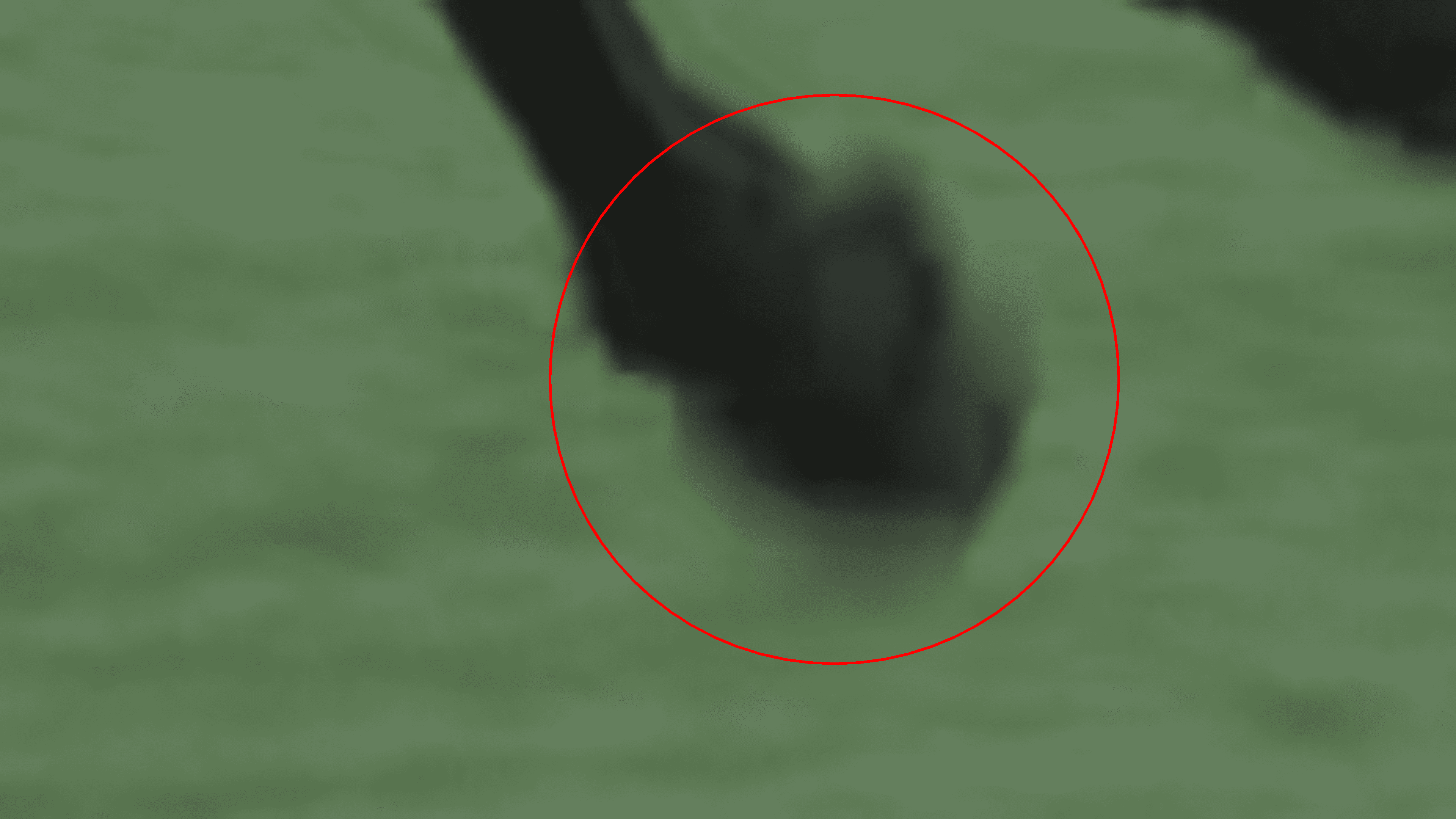}&
          \includegraphics[width=0.1\linewidth]{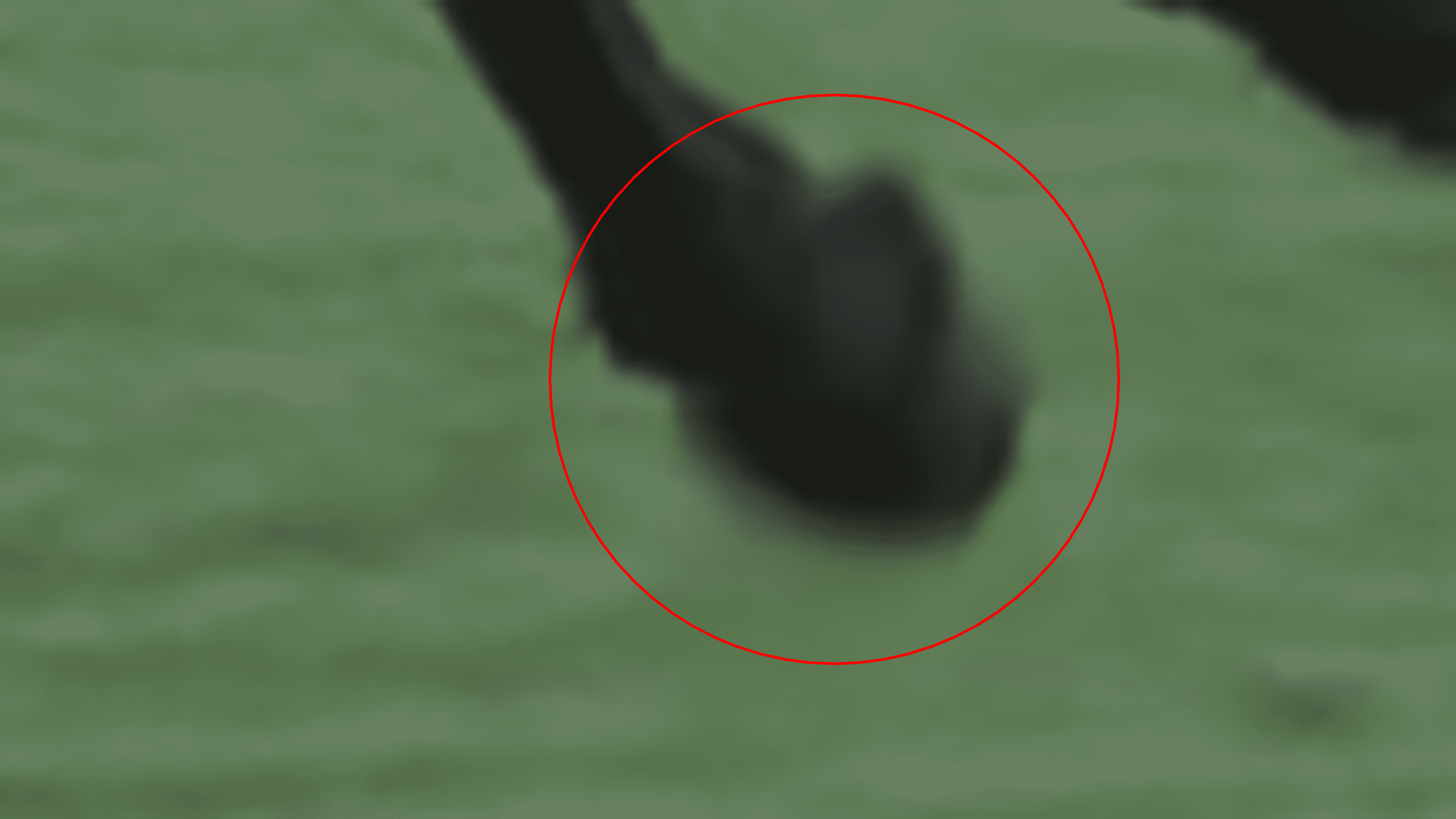}&
          \includegraphics[width=0.1\linewidth]{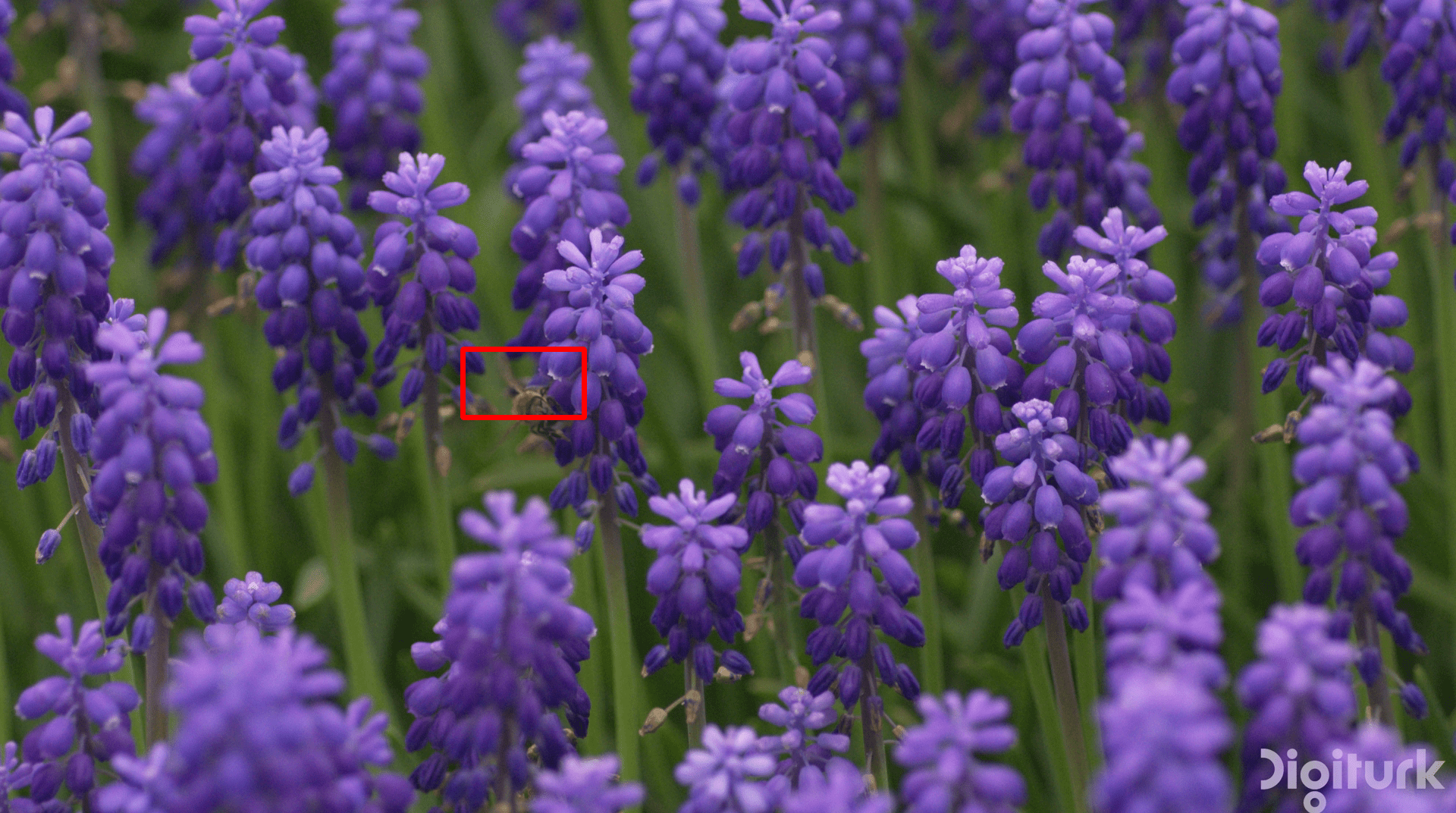}&
          \includegraphics[width=0.1\linewidth]{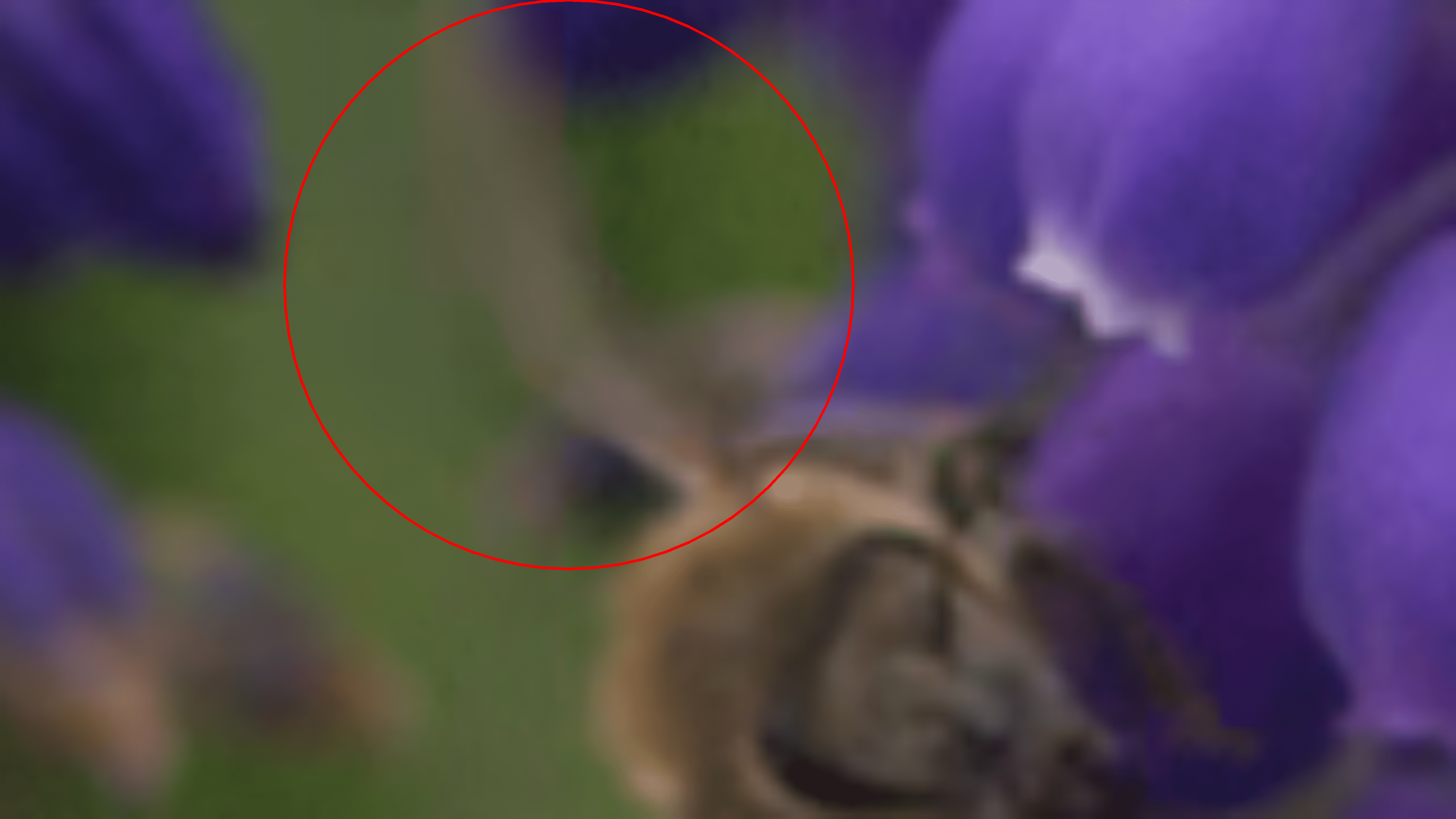}&
          \includegraphics[width=0.1\linewidth]{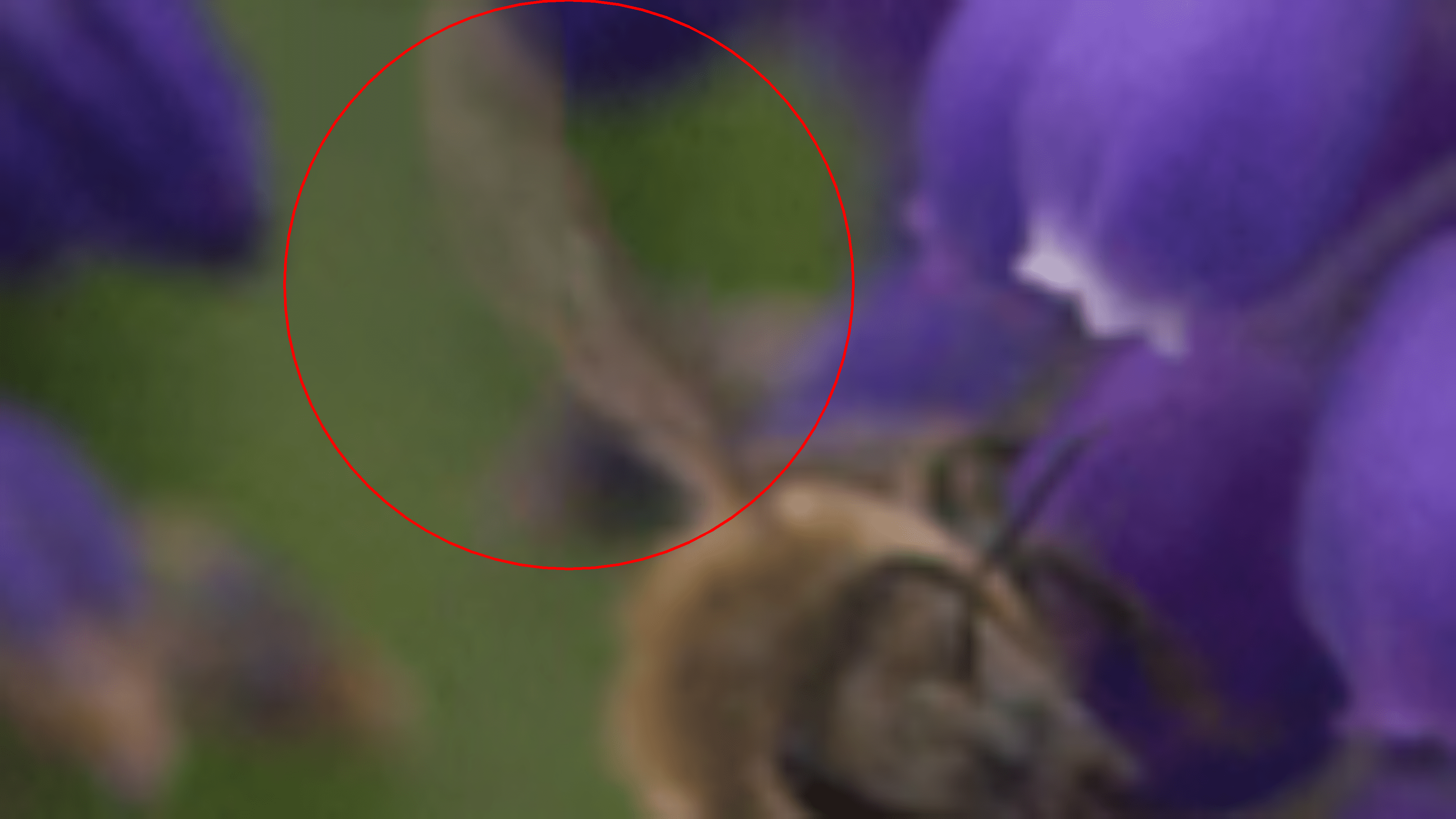}\\
          \tiny{\texttt{UVG [Jockey]}} & \tiny{\texttt{Gao + DCVC}} & \tiny{\texttt{Ours}} & \tiny{\texttt{UVG [HoneyBee]}} & \tiny{\texttt{Gao + DCVC}} & \texttt{Ours}
        
\end{tabular}}
\end{center}
\vspace{-5mm}
\caption{\small Qualitative analysis on real-world videos. Note that PSNR is not possible to compute as sharp GT is not available.}
\label{tbl:reb_fig}
\vspace{-5mm}
\end{figure*}
%%%%%%%%%%%%%%%%%%%%%%%%%%%%%%%%%%%%%%%%%%%%%%%
\vspace{-3mm}
\paragraph{Bit-rate and Distortion.} \Fref{fig:exp_results}a depicts the rate-distortion curves of our method and \textit{off-the-shelf} cascade models on Adobe240 \cite{su2017deep}, GOPRO \cite{nah2017deep} and REDS \cite{Nah_2019_CVPR_Workshops_REDS} test sets. As can be inferred from the figure, our approach performs favorably against all possible baselines on PSNR metric. For instance, given a blurred input and a budget of 0.3 Bpp, our approach can reconstruct sharp frames at a quality of 29.52 dB on average, while the second best performing method (Gao \cite{gao2019dynamic} + DCVC \cite{li2021deep}) reconstructs frames at a quality of 27.68 dB. The main reason behind such a performance gap is due to error propagation in \textit{off-the-shelf} cascade models, where blur/compression artifacts propagate from the first stage to the second stage, degrading the overall performance. Moreover, as each network is optimized for the individual task, na\"ively cascading them for the joint task often gives sub-optimal results as observed in other works \cite{shen2020blurry,zhang2020video}. It can also be inferred from \Fref{fig:exp_results}a that \texttt{D} + \texttt{C} models perform significantly better than their \texttt{C} + \texttt{D} counterparts. This is mainly because deblurring networks~\cite{gao2019dynamic} perform poorly in the presence of both blur and compression artifacts. 

One way to address the limitations of \textit{off-the-shelf} cascade models would be to end-to-end optimize trainable cascade models. In \Fref{fig:exp_results}b, we study the  \textit{joint training} scheme, where cascade models are trained as one network with the standard rate-distortion loss ($\calL_{\texttt{Joint}}$). As can be seen from \Fref{fig:exp_results}b, the \textit{joint training} scheme results in performance worse than \textit{off-the-shelf} models. This is mainly because of the inherent trade-off between the two tasks as discussed in \Sref{sec:intro}. Our experimental analysis reveals that, for \texttt{D} + \texttt{C} models, the deblurring network basically collapses to an identity function. This is intuitive because it is easier to compress a blurry input compared to its sharper equivalent since the latter one will have more information to encode. As a consequence, the compression network alone has to reconstruct the input video while reducing its temporal redundancies but deblurring its contents at the same time, which are incompatible. Similarly, we observed that \texttt{C} + \texttt{D} models incur a heavy encoding cost as shown in \Fref{fig:exp_results}b. This is because the compression network diverges to preserving as much information as possible for the deblurring network to enhance. These results show that the trade-off balance always tips to the second task when cascade models are trained in the \textit{joint training} scheme, resulting in a subpar overall performance on the joint task.

%%%%%%%%%%%%%%%%%%%%%%%%%%%%%%%%%%%%%%%%%%%%%%%%%%%%%%%%%%%%%%%%%%%%%%%%%%%%
\begin{figure*}[!t]
    \centering
    \begin{subfigure}[]{0.24\textwidth}
    \centering
    \includegraphics[width = \textwidth, trim={0.35cm 0.25cm 0.35cm 0.35cm},clip]{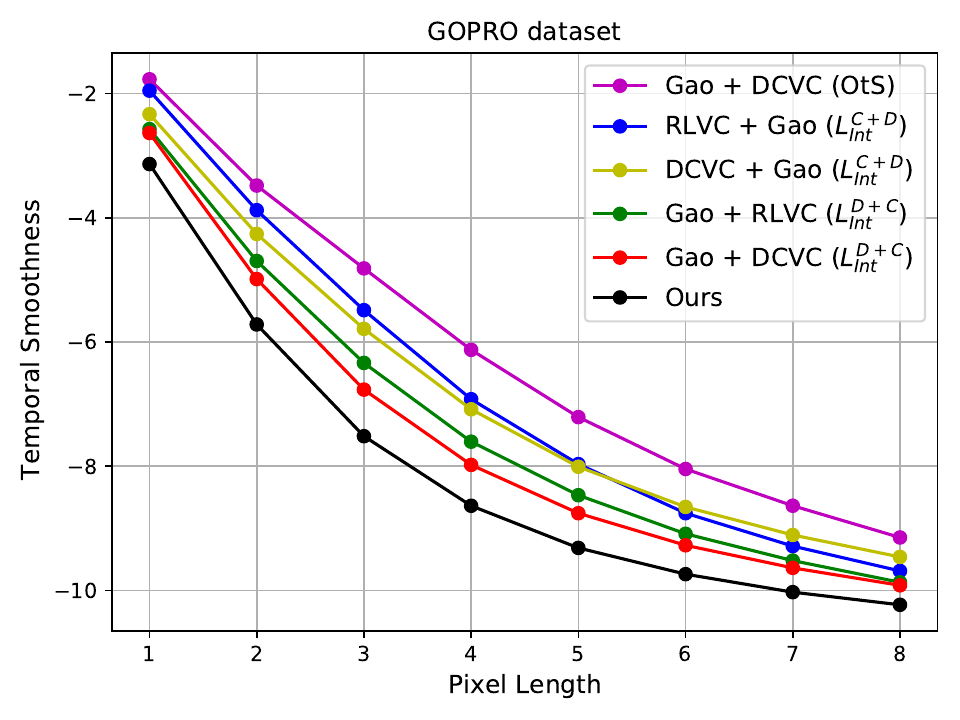}
    \caption{}
    \end{subfigure}
    \hfill
    \begin{subfigure}[]{0.24\textwidth}
    \centering
    \includegraphics[width = \textwidth, trim={0.83cm 0.25cm 1.60cm 0.8cm},clip]{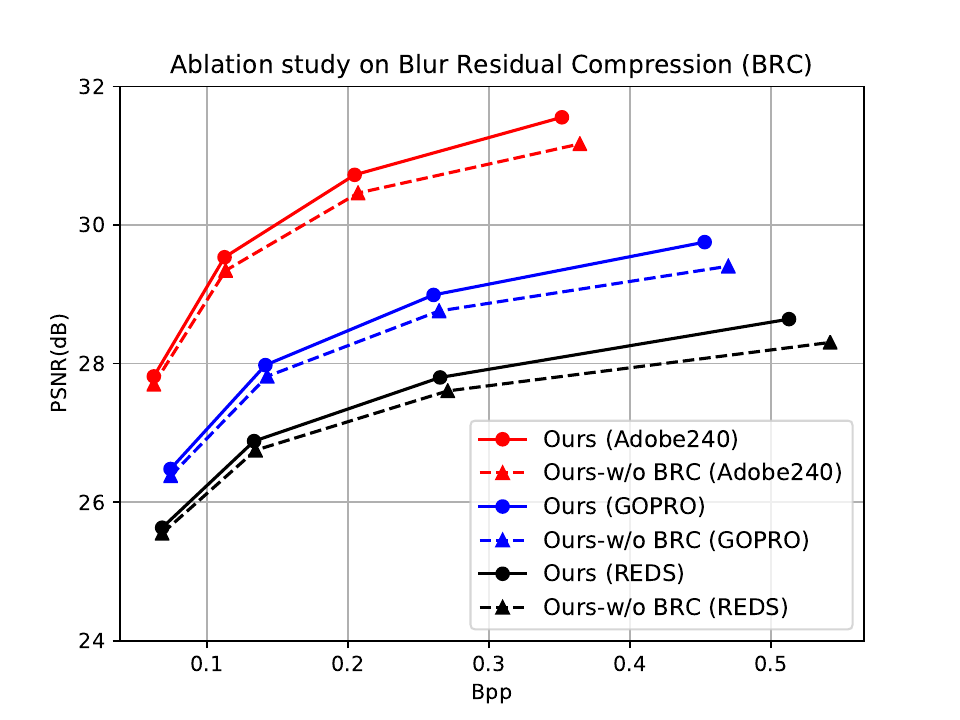}
    \caption{}
    \end{subfigure}
    \hfill
    \begin{subfigure}[]{0.24\textwidth}
    \centering
    \includegraphics[width = \textwidth, trim={0.83cm 0.25cm 1.50cm 0.8cm},clip]{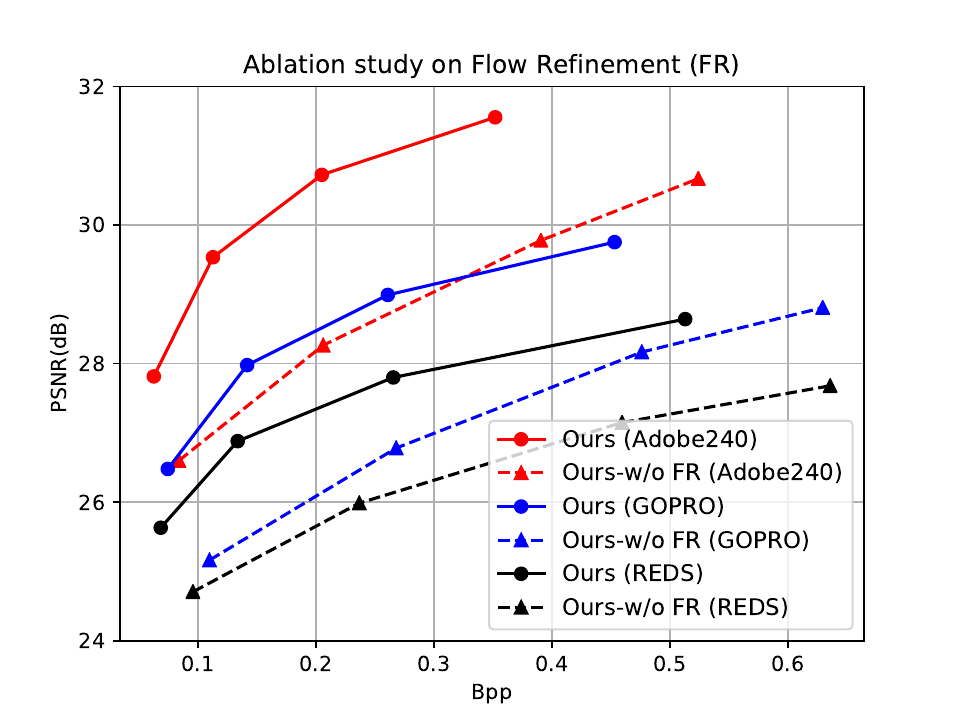}
    \caption{}
    \end{subfigure}
    \hfill
    \begin{subfigure}[]{0.24\textwidth}
    \centering
    \includegraphics[width = \textwidth, trim={0.83cm 0.25cm 1.60cm 0.8cm},clip]{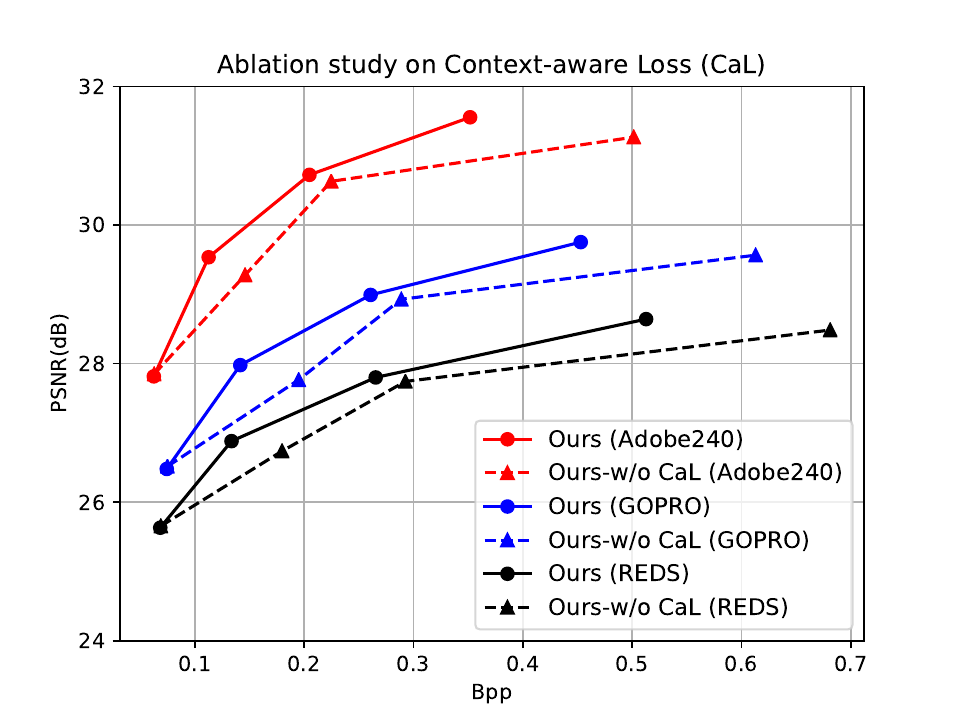}
    \caption{}
    \end{subfigure}
    \vspace{-3mm}
    \caption{\small (\textbf{a}) Temporal smoothness comparison, (\textbf{b}) - (\textbf{d}) Ablation studies on blur residual compression, flow refinement and $\calL_\texttt{CaL}$}
    \label{fig:ablation}
    \vspace{-4mm}
\end{figure*}
%%%%%%%%%%%%%%%%%%%%%%%%%%%%%%%%%%%%%%%%%%%%%%%%%%%%%%%%%%%%%%%%%%%%%%%%

To create a stronger baseline, we optimize the cascade models using a training scheme analogous to our proposed training strategy, \ie~\textit{intermediate supervision} in \Eref{eq : optim2} and \Eref{eq : optim2_2}, where we introduce additional intermediate supervision between the cascaded components. As can be seen from \Fref{fig:exp_results}c, the proposed training strategy significantly improves the performance of cascade models. It can also be inferred from \Fref{fig:exp_results}c that \texttt{D} + \texttt{C} methods generally outperform their corresponding \texttt{C} + \texttt{D} counterparts despite the fact that \texttt{C} + \texttt{D} take much fewer bits given that the compression network encodes the blurry input in this setting. This is mainly because the deblurring network fails to blindly enhance the compressed output due to the high ill-posedness of the task. Enhancing the quality of compressed videos is a very challenging problem even for sharp videos \cite{yang2018multi}, let alone for \textit{blurred} and \textit{compressed} inputs.

In comparison to the strongest \texttt{D + C} models, our network still gives a notably better result. To further analyze the results, we compute the quality gain between the highest and smallest distortions for the curves in \Fref{fig:exp_results}c over the corresponding increase in bit-rate cost. In this metric, our approach achieves 10.80 dB/Bpp on average whereas Gao \cite{gao2019dynamic} + DCVC~\cite{li2021deep} could only achieve 7.86 dB/Bpp. The performance gain of our method compared to cascade \texttt{D + C} models trained under the same setting can be attributed to the motion refinement step in \Sref{sec:step2} which is crucial for optimal compression performance in a blurry scenario as opposed to fine-tuning a pretrained flow network (see \Sref{sec:ablation}). In \Fref{tbl:reb_fig}, we show qualitative results on blurry scenes from real-world videos. As shown in the figure, our model gives a visually better output at a lower encoding cost compared to the strongest baseline. 

\subsection{Experimental Analyses}
\paragraph{Preserving the Input Prior.} Although motion blur is mostly considered an unwanted artifact, it can be sometimes useful to add realism to a scene. Taking that into consideration, we experiment with the idea of having the option to preserve the blur when performing compression. We achieve this during \textit{inference} time by simply using input $B_t$ (instead of the enhanced frame $\hat{x}_t$) to compute the frame residual $r_t$ in \Sref{sec:step3}, \ie~$r_t = B_t -\overline{x}_t$. The remaining steps follow accordingly.This mechanism is an inverse process of the visual enhancement in \Sref{sec:step1} where $r_t$ is equivalent to the blur residual $b_t$. As previous works \cite{Agustsson_2020_CVPR,li2021deep,yang2020learningb} typically output a blurry video given a blurry input, we use them as a baseline to evaluate our work on preserving a prior. As can be inferred from \Fref{fig:exp_results}d (\textit{top}), our approach (with the simple fix) performs better than state-of-the-art VC approaches in maintaining the input blur when compressing a blurry video. This result further highlights the flexibility of our proposed framework to be easily adapted to diverse conditions without the need for re-training. 

The notion of preserving  a prior also applies when the input video is distinctively sharp with predominantly static contents. To compare our approach and previous works in this scenario, we use the UVG \cite{mercat2020uvg} and MCL-JCV~\cite{wang2016mcl} datasets. As can be seen in \Fref{fig:exp_results}d (\textit{middle} and \textit{bottom}), our model trained in a blurry setup gives a very competitive performance compared to the  state-of-the-art models \cite{lu2019dvc,yang2020learninga,yang2020learningb} optimized on a sharp dataset.
\vspace{-3mm}
\paragraph{Complexity Analysis.} In \Tref{tbl:quant_true}, we analyze the runtime and parameter size of our model and the competing baselines. As shown in the table, our approach takes $0.67$ seconds to reconstruct an input frame of size $1280 \times 720$px while the best-performing cascade models, \ie~Gao \cite{gao2019dynamic} + DCVC~\cite{li2021deep} and Gao~\cite{gao2019dynamic} + RLVC~\cite{yang2020learninga}, take $1.97$ and $1.98$ seconds, respectively.
\vspace{-3mm}
\paragraph{Temporal Smoothness.} To evaluate the smoothness of videos decoded by our method and competing baselines, we adopt a flow-based smoothness metric~\cite{Jin_2019_CVPR,shen2020blurry}. We first compute the second-order differential flow between the decoded video and its GT counterpart using 3 consecutive frames at a time, \ie~$df = (f_{x_2 \rightarrow x_1}-f_{x_1 \rightarrow x_0})-(f_{X_2 \rightarrow X_1} - f_{X_1 \rightarrow X_0})$, where $x_0, x_1, x_2$, and $X_0, X_1, X_2$ are frame sequences in the decoded and GT videos, respectively. The temporal smoothness metric $\calT(l)$ is then defined as function of the pixel error length $l$,
\vspace{-2mm}
\begin{equation}
    \calT(l) = \log \sum_{v \in df} \mathbf{1}_{[l, l+1)}(\|v\|_2) - \log |df|
\vspace{-2mm}
\end{equation}
where $v$ denotes a vector of matrix $df$, $|\cdot|$ represents the size of a matrix and the indicator function $\mathbf{1}_S (x)$ equals to 1 if $x$ belongs to set $S$. The lower value of $\calT(l)$ indicates better temporal stability. As can be seen from \Fref{fig:ablation}(a), \textit{off-the-shelf} cascade models perform worse on the temporal smoothness metric as they are prone to error propagation. Compared to the strongest baseline (Gao \cite{gao2019dynamic} + DCVC \cite{li2021deep} trained with $\calL_\texttt{Int}^{\texttt{D} + \texttt{C}}$), our approach decodes a temporally smoother video.

\section{Ablation Study}
\label{sec:ablation}

In this section, we perform extensive experiments to show the effectiveness of our designed modules. First, we study the importance of blur residual compression by training a network without the $b_t$ auto-encoder, where the enhanced frame $\hat{x}_t$ is obtained directly by adding $b_t$ to the input $B_t$, \ie~$\hat{x}_t = B_t + b_t$. It can be inferred from \Fref{fig:ablation}(b) that blur residual compression consistently leads to better performance. As mentioned previously, compressing the blur residual information plays a significant role in maintaining the balance between visual enhancement and compression for optimal overall performance. 

We analyze the importance of the motion refinement step by training our network without \texttt{FRNet}, \ie~we fine-tuned the optical flow network using a warp loss. As can be seen from \Fref{fig:ablation}(c), directly fine-tuning the optical flow network in a blurry scenario results in significantly worse performance compared to a network trained with flow refinement. This is mainly because of erroneous initial flow estimation since the pixels in the input frames are corrupted by blur and a blind warp loss in \Eref{eqn: warp} can not address this limitation (refer to \Sref{sec:step2}). To demonstrate the benefit of the proposed context-aware loss ($\calL_\texttt{CaL}$), we train the \texttt{FRNet} using the warp loss ($\calL_\texttt{warp}$ in \Eref{eqn: warp} using $f_{t \rightarrow t-1}$) and compare the results with a network trained using $\calL_\texttt{CaL}$. As shown in \Fref{fig:ablation}(d) and \Fref{fig:qual_vis}, attending to the visual enhancement using $\calE$ leads to better motion estimation and notably superior compression performance.

%%%%%%%%%%%%%%%%%%%%%%%%%%%%%%%%%%%%%%%%%%%%%%%%%%%%%%%%%%%%%
\begin{table}[!t]
\setlength{\tabcolsep}{4.5pt}
\renewcommand{\arraystretch}{1}
\caption{\small Evaluation on time complexity and parameter size}
\vspace{-2mm}
\begin{adjustbox}{width=\linewidth}
\begin{tabular}{l|c|c|c|c|c}
\toprule
Method &  {\makecell{Gao~\cite{gao2019dynamic} \\ + \\  H.265~\cite{sullivan2012overview}}}  &  {\makecell{Gao~\cite{gao2019dynamic} \\ + \\  SFF~\cite{Agustsson_2020_CVPR}}}  &   {\makecell{Gao~\cite{gao2019dynamic} \\ + \\  RLVC~\cite{yang2020learninga}}}  &  {\makecell{Gao~\cite{gao2019dynamic} \\ + \\  DCVC~\cite{li2021deep}}}  &   {\makecell{Ours \\ \\ }} \\ \midrule 
Parameters ($\times 10^ 6$) & \textbf{3.87} & 38.08 & 20.56 & 11.81 & 12.74 \\ \midrule
Runtime (s) & 1.52 & 2.35 & 1.98 & 1.97 & \textbf{0.67}\\ \bottomrule
\end{tabular}
\end{adjustbox}
\label{tbl:quant_true}
\vspace{-5mm}
\end{table}
%%%%%%%%%%%%%%%%%%%%%%%%%%%%%%%%%%%%%%%%%%%%%%%%%%%%%%%%%%%%%
\vspace{-1mm}
\section{Conclusion}
\label{sec:conclusion}

In this work, we tackle the video compression problem in a general situation, where unwanted blurs may be present in videos. We design various cascade models as baselines and analyze their limitations. To overcome these limitations, we propose a novel framework that can be efficiently optimized in an end-to-end manner. We have demonstrated the effectiveness and flexibility of our approach through extensive analyses on different datasets. However, there remain a few limitations. In extreme cases, where videos are severely blurred or temporally undersampled, we experimentally observed that our enhancement module \texttt{VENet} fails, therefore compromising the overall compression performance.

\paragraph{Acknowledgment} This work was supported by the National Research Foundation of Korea (NRF) grant funded by the Korea government (MSIT, No. RS-2023-00212845).

% {

% }
%%%%%%%%% REFERENCES
{\small
\bibliographystyle{ieee_fullname}
\bibliography{egbib}
}

\end{document}